\documentclass[12pt]{article}
\pdfoutput=1

\newlength{\abstractwidth}
\usepackage{amsmath}
\usepackage{amsfonts}
\usepackage{amssymb}
\usepackage{latexsym}

\usepackage{inputenc}
\usepackage{epsf}
\usepackage{color}
\usepackage{graphicx}
\usepackage{tikz}
\usepackage{dsfont}

\usetikzlibrary{arrows,shapes,positioning}
\usetikzlibrary{decorations.markings}
\usepackage[rightcaption]{sidecap}
\tikzstyle arrowstyle=[scale=1]
\tikzstyle directed=[postaction={decorate,decoration={markings,
    mark=at position .65 with {\arrow[arrowstyle]{stealth}}}}]
\tikzstyle reverse directed=[postaction={decorate,decoration={markings,
    mark=at position .65 with {\arrowreversed[arrowstyle]{stealth};}}}]
\usetikzlibrary{positioning}

\usepackage{hyperref}
\definecolor{darkred}{rgb}{0.8,0.1,0.1}
\hypersetup{colorlinks=true, linkcolor=darkred, citecolor=blue, linktoc=page}

\renewcommand{\thefootnote}{\fnsymbol{footnote}}
\renewcommand{\thanks}[1]{\footnote{#1}}
\newcommand{\starttext}{
\setcounter{footnote}{0}
\renewcommand{\thefootnote}{\arabic{footnote}}}

\newcommand{\bea}{\begin{eqnarray}}
\newcommand{\eea}{\end{eqnarray}}
\newcommand{\be}{\begin{eqnarray}}
\newcommand{\ee}{\end{eqnarray}}
\newcommand{\bal}{\begin{align}}
\newcommand{\eal}{\end{align}}
\newcommand{\bma}{\begin{matrix}}
\newcommand{\ema}{\end{matrix}}

\DeclareMathOperator{\vol}{vol}

\def\cA{{\cal A}}
\def\cB{{\cal B}}
\def\cC{{\cal C}}

\def\cG{{\cal G}}

\def\cM{{\cal M}}

\def\cO{{\cal O}}

\def\cW{{\cal W}}

\def\ZZ{{\mathbb Z}}
\def\RR{{\mathbb R}}

\def\CC{{\mathbb C}}

\def\Re{{\rm Re \,}}
\def\Im{{\rm Im \,}}

\def\half{{1\over 2}}
\def\p{\partial}

\def\a{\alpha}
\def\b{\beta}
\def\g{\gamma}

\def\ep{\varepsilon}

\def\ZZ{{\mathds Z}}
\def\RR{{\mathds R}}

\def\CC{{\mathds C}}

\def\PP{{\mathds P}}

\def\no{\nonumber}
\def\sm{\smallskip}

\usepackage{bbold}

\usepackage{verbatim} % comment

\usepackage{titling}
\usepackage{cite}

\numberwithin{equation}{section} %equations numbered by section (amsmath)
\begin{document}
\starttext
\setcounter{footnote}{0}

\title{Global half-BPS $AdS_{2}\times S^{6}$ solutions in Type~IIB}
\author{{David Corbino, Eric D'Hoker, Justin Kaidi, Christoph F.~Uhlemann} \\[4mm]
{\sl  Mani L. Bhaumik Institute for Theoretical Physics}\\
{\sl Department of Physics and Astronomy} \\
{\sl University of California, Los Angeles, CA 90095, USA}\\
{\tt \small corbino, dhoker, jkaidi, uhlemann@physics.ucla.edu}}

\date{\today}

%%%
\begin{titlingpage}
\maketitle

\begin{abstract}
We investigate half-BPS Type IIB supergravity solutions with spacetime geometry $AdS_2\times S^6$ warped over a Riemann surface $\Sigma$.  The general local solution was obtained in earlier work in terms of two holomorphic functions $\mathcal A_\pm$ on $\Sigma$. In the first part of this paper we seek global solutions corresponding to the near-horizon behavior of $(p,q)$-string junctions. We identify the type of singularity in $\mathcal A_\pm$ needed at the boundary of $\Sigma$ to match the solutions locally onto the classic  $(p,q)$-string solution.  We construct solutions with multiple $(p,q)$-strings, but the existence of geodesically complete solutions remains unsettled. In a second part we  construct multi-parameter families of non-compact globally regular and geodesically complete solutions with asymptotic regions and an $AdS_2$ throat which caps off smoothly.
\end{abstract}

\end{titlingpage}
%%%
\vfill\eject
%%%
\setcounter{tocdepth}{2} 
\tableofcontents
\vfill\eject

\baselineskip=15pt
\setcounter{equation}{0}
\setcounter{footnote}{0}

%%%%%%%%%%%%%%%%%%%%%%%%%%%%%%%%%%%%%%%%%%%%%%%%%%
%%%%%%%%%%%%%%%%%%%%%%%%%%%%%%%%%%%%%%%%%%%%%%%%%%
\section{Introduction}
%%%%%%%%%%%%%%%%%%%%%%%%%%%%%%%%%%%%%%%%%%%%%%%%%%
%%%%%%%%%%%%%%%%%%%%%%%%%%%%%%%%%%%%%%%%%%%%%%%%%%

The classification of supergravity solutions containing a factor of anti-de Sitter space is a topic of long-standing interest in  the construction of holographic duals to conformal field theories. With the maximal number of 32 supersymmetries, the classic supergravity solutions $AdS_5 \times S^5$, $AdS_4 \times S^7$, and $AdS_7 \times S^4$ - invariant respectively under the Lie superalgebras $PSU(2,2|4), OSp(8|4;\RR)$, and $OSp(8^*|4)$ - served as prototypes for the AdS/CFT correspondence  \cite{Maldacena:1997re}. A classification of sub-superalgebras  of these maximally supersymmetric algebras containing 16 supersymmetries was developed in \cite{DHoker:2008wvd}, and many of the predicted solutions have since been constructed.

\sm

Motivated by brane considerations \cite{Brandhuber:1999np}, progress was recently made on the construction of supergravity solutions \cite{Apruzzi:2014qva,Kim:2015hya,D'Hoker:2016rdq} which are holographic duals to five-dimensional superconformal field theories. The superconformal algebra in five dimensions is unique and given by a particular real form of the exceptional Lie superalgebra $F(4)$ with maximal bosonic subalgebra $SO(2,5) \oplus SO(3)$. This is not a sub-superalgebra of any of the three maximal superalgebras. The corresponding Type IIB supergravity solutions have a spacetime of the form $AdS_6 \times S^2$ warped over a Riemann surface with boundary $\Sigma$, and are specified in terms of two locally holomorphic functions $\cA_\pm$ on $\Sigma$. Globally regular and geodesically complete solutions sourced by the charges $p,q$  of the complex  three-form field strength of Type IIB were shown to provide fully back-reacted geometries for the near-horizon region of general $(p,q)$ five-brane webs  \cite{DHoker:2016ysh, DHoker:2017mds,DHoker:2017zwj}. 

\sm

By double analytic continuation from Minkowski signature $AdS_6$ and $AdS_2$  to Euclidean signature $S^6$ and $S^2$, the existence of half-BPS $AdS_6 \times S^2$ solutions suggests the existence of half-BPS solutions with spacetime $AdS_2 \times S^6$. Indeed, general local solutions of this form were constructed in \cite{Corbino:2017tfl}. As in the case of $AdS_6 \times S^2$, the solutions are given  in terms of two locally holomorphic functions $\cA_\pm$ on a Riemann surface $\Sigma$ \cite{Corbino:2017tfl}. Their symmetry group is given by a different real form of the Lie superalgebra $F(4)$, with maximal bosonic superalgebra $SO(1,2) \oplus SO(7)$, which is again not a sub-superalgebra of any of the three maximal supersymmetric algebras. 

\sm

The purpose of this  paper is two-fold. In a first part, we shall investigate the existence of global $AdS_2 \times S^6$ solutions sourced by seven-form  charges $p,q$, which are naturally associated with $(p,q)$-strings. We shall examine the emergence of $(p,q)$-string web solutions  \cite{Sen:1997xi,Bergman:1998gs,Lunin:2008tf} in the near-horizon limit. Although the supergravity fields of the $AdS_2 \times S^6$ solutions differ from those of the $AdS_{6}\times S^{2}$ solutions merely by certain sign reversals, these simple differences make the construction of globally regular $AdS_2\times S^6$ solutions intricate and technically difficult. While we shall succeed in producing solutions with multiple $(p,q)$-strings in the near-horizon limit, the geodesic completeness of such solutions remains unsettled. In a second part we shall study solutions independently from any string junction interpretation.  We present multi-parameter families of globally regular solutions, which have asymptotic regions where spacetime decompactifies and an $AdS_2\times S^6$ ``throat'' that caps off smoothly. 

\sm

While the superconformal algebra $F(4)$ is unique in five-dimensions, there exist four distinct superconformal algebras with 16 supersymmetries in the one-dimensional theories  holographically dual to an $AdS_2$ factor \cite{VanProeyen:1986me}. As mentioned above, the real form of $F(4)$ with maximal bosonic subalgebra $SO(2,1) \oplus SO(7)$ is studied in this paper. $OSp(4^*|4)$ enters the holographic dual to a Wilson line  constructed in \cite{DHoker:2007mci}, while holographic duals to the remaining cases $SU(1,1|4)$ and $OSp(8|2,\RR)$ are currently being investigated.

\sm

Supersymmetric $AdS_2$ solutions have been studied in a wide variety of contexts, including $AdS_2/CFT_1$ dualities \cite{Kim:2006qu,Hartman:2008dq}, relations with the SYK model  \cite{Almheiri:2014cka,Jensen:2016pah,Engelsoy:2016xyb,Maldacena:2016upp},  for compactifications of higher-dimensional field theories \cite{Suh:2018tul,Hosseini:2018uzp,Hosseini:2018usu,Suh:2018szn}, and as black hole near-horizon geometries. Further studies of $AdS_2$ solutions in Type IIA and M-theory can be found in \cite{Massar:1999sb,Cvetic:2000cj,Imamura:2001cr,Donos:2008ug,Dibitetto:2018gbk,Dibitetto:2018gtk}.

\sm

The remainder of the paper is organized as follows. In sec.~\ref{sec:review} we review the local $AdS_2\times S^6$ solutions of \cite{Corbino:2017tfl}. In sec.~\ref{sec:string-ansatz} we construct an Ansatz suitable for string junction solutions and show that they match locally to the classic $(p,q)$-string solution. In sec.~\ref{sec:global-sol} we construct multi-parameter families of globally regular solutions. In sec.~ \ref{app:T-dual}, we briefly study the T-duals of the D0-F1-D8 system in massive Type IIA supergravity \cite{Dibitetto:2018gbk}. For the case of $AdS_6 \times S^2$, a subset of the solutions were found to correspond to Abelian and non-Abelian T-duals of the D4-D8 system \cite{D'Hoker:2016rdq,Hong:2018amk,Lozano:2018pcp}. On the other hand, we will find that the T-duals of the analogous D0-F1-D8 system are not of the form $AdS_2 \times S^6$. We conclude with a discussion in sec.~\ref{sec:discussion}.

%%%%%%%%%%%%%%%%%%%%%%%%%%%%%%%%%%%%%%%%%%%%%%%%%%
%%%%%%%%%%%%%%%%%%%%%%%%%%%%%%%%%%%%%%%%%%%%%%%%%%
\section{Local solution and regularity conditions}
\label{sec:review}
%%%%%%%%%%%%%%%%%%%%%%%%%%%%%%%%%%%%%%%%%%%%%%%%%%
%%%%%%%%%%%%%%%%%%%%%%%%%%%%%%%%%%%%%%%%%%%%%%%%%%

We summarize the local form of Type IIB supergravity solutions with 16 supersymmetries and spacetime of the form $AdS_{2}\times S^{6}$ warped over a Riemann surface $\Sigma$ obtained in \cite{Corbino:2017tfl}, and  discuss the conditions for physical  positivity and regularity of the supergravity fields. The local solutions are invariant under the real form of the exceptional Lie superalgebra $F(4)$ which has maximal bosonic subalgebra $SO(1,2) \oplus SO(7)$.

\subsection{Supergravity fields}

Invariance under $SO(1,2) \oplus SO(7)$ dictates the general form of the supergravity fields of the solutions. All Fermi fields vanish and the spacetime metric takes the form, 
\begin{align}
\label{eq:metric}
ds^{2} = f_{2}^{2}ds_{AdS_{2}}^{2} + f_{6}^{2}ds_{S^{6}}^{2} + 4\rho^2 |dw|^{2}
\end{align}
The five-form field strength vanishes $F_{(5)}=0$ and the three-form field strength $F_{(3)}$ and its Poincar\'e dual $F_{(7)}$ are given by \cite{Corbino:2017tfl},
\begin{align}
F_{(3)} & = d C_{(2)} & C_{(2)} & = \mathcal{C} ~\textrm{vol}_{AdS_{2}} 
\no \\ 
F_{(7)} & = d C_{(6)} & C_{(6)} & = \mathcal{M} ~\textrm{vol}_{S^{6}}
\end{align}
Throughout, $w$ is a local complex coordinate on $\Sigma$ while $f_2$, $f_6$, and $\rho$ are real-valued functions on $\Sigma$. The fields $\cC$, $\cM$, and the axion-dilaton $B=(1+i\tau)/(1-i\tau)$ are complex-valued functions on $\Sigma$. The line elements $ds_{AdS_{2}}^{2}$,$ds_{S^{6}}^{2}$ and the volume forms $\textrm{vol}_{AdS_{2}}$,$\textrm{vol}_{S^{6}}$ are for maximally symmetric $AdS_2$ and $S^6$ with unit radius.

\sm

The solutions are parametrized by two locally holomorphic functions $\cA_\pm$ and expressed conveniently in terms of the composite quantities $\kappa ^2$, $\cG$, and $T$ given in terms of $\cA_\pm$ by,
\begin{align}
\label{eq:comp} %(\ref{eq:comp})
\kappa^{2} & = -|\partial_{w}\mathcal{A}_{+}|^{2} + |\partial_{w}\mathcal{A}_{-}|^{2} & \partial_{w}\mathcal{B} & = \mathcal{A}_{+}\partial_{w}\mathcal{A}_{-} - \mathcal{A}_{-}\partial_{w}\mathcal{A}_{+} 
\no \\
\mathcal{G} & = |\mathcal{A}_{+}|^{2} - |\mathcal{A}_{-}|^{2} + \mathcal{B} + \bar{\mathcal{B}} & 
T&=\frac{1-R}{1+R}=
 \left (1 + \frac{2|\partial_{w}\mathcal{G}|^{2}}{3\kappa^{2}\mathcal{G}}\right )^\half
\end{align}
Note that $\kappa^2=-\partial_w\partial_{\bar w}\cG$. By construction \cite{Corbino:2017tfl}, the functions $\kappa^2, \cG,$ and $R$ are real. Furthermore, $R$ is non-negative so that $T$ is real and satisfies $T\in[-1,1]$. In terms of these composites, the metric functions are given by, 
\begin{align}
\label{f2f6rho}
f_{2}^{2} & = \frac{1}{9}\left ( \frac{-6\mathcal{G}}{T^3} \right )^\half & 
f_{6}^{2} & = \left ( -6\mathcal{G} T \right )^\half &
 \rho^{2} & = \kappa^{2}\left ( \frac{T}{-6\mathcal{G}} \right )^\half
\end{align}
The functions $\cC$ and $\cM$ parametrizing the two- and six-form potentials are given by,
\begin{align}
\mathcal{C}  &= -\frac{2i }{3}\left ( \frac{U}{3T^{2}} - \bar{\mathcal{A}}_{-} - \mathcal{A}_{+} \right )
\no\\
  \cM  &=
  80 (\cW_++\bar\cW_-)
 -12\cG\,U
 +20(\cA_++\bar\cA_-) \Big (2|\cA_+|^2-2|\cA_-|^2-3\cG \Big )
\label{eq:cC-cM}
\end{align}
where $U$ and $\cW_\pm$ are defined by,
\bea
\kappa ^2 U=\overline{\partial_{w}\mathcal{G}}\partial_{w}\mathcal{A}_{+} + \partial_{w}\mathcal{G}\overline{\partial_{w}{\mathcal{A}}_{-}}
\hskip 1in 
 \p_w \cW_\pm = \cA_\pm \p_w \cB 
\eea
The axion-dilaton scalar field  is given by,
\begin{align}
B & = - \frac{\partial_{w}\mathcal{A}_{+}\partial_{\bar{w}}\mathcal{G} + R\, \partial_{\bar{w}}\bar{\mathcal{A}}_{-}\partial_{w}\mathcal{G}}{R \, \partial_{\bar{w}}\bar{\mathcal{A}}_{+}\partial_{w}\mathcal{G} + \partial_{w}\mathcal{A}_{-}\partial_{\bar{w}}\mathcal{G}}
\end{align}
The global $SU(1,1)$ symmetry transformations of the Type IIB supergravity fields are induced by the following  transformations of $\mathcal{A}_{\pm}$ under the group $SU(1,1)\otimes \mathds{C}$ ,
\bea\label{eq:SU11}
\mathcal{A}_{+} & \to & \mathcal{A}'_{+}  =  +u\mathcal{A}_{+} - v\mathcal{A}_{-} + a 
\no \\
\mathcal{A}_{-} & \to & \mathcal{A}'_{-}  =  -\bar{v}\mathcal{A}_{+} + \bar{u}\mathcal{A}_{-} + \bar a
\eea
where $SU(1,1)$ is parametrized by $u, v \in \mathds{C}$ with $|u|^{2} - |v|^{2} = 1$ and the complex
shift parameter $a$ has the effect of producing gauge transformations in $\cC$ and $\cM$ only.

\subsection{Positivity and regularity conditions}
\label{sec:bc-review}

Minkowski signature of the ten-dimensional spacetime metric  imposes the positivity conditions $f_2^2, f_6^2, \rho^2>0$ which require $\kappa ^2 >0$ and $\cG T <0$, assuming that all square roots of positive real arguments in (\ref{f2f6rho}) are taken to be positive (see sec.~7 of \cite{Corbino:2017tfl}). Without loss of generality, we may choose the branch $T>0$ for the square root in (\ref{eq:comp}), so that $0<R<1$. As a result, the positivity conditions become,
\bea \label{eq:5_4h}
\kappa ^2 >0 \hskip 1in \cG<0 \hskip 1in 0<R<1
\eea
Regularity of the supergravity fields of the solutions in the interior of $\Sigma$ requires that the inequalities of (\ref{eq:5_4h}) be obeyed strictly. If $\Sigma$ has a non-empty boundary $\p \Sigma$, then geodesic completeness of the solutions requires that the six-sphere shrinks to zero size $f_6 \to 0$ at the boundary, while the radius of $AdS_2$ remains finite. Since we have $f_6^2/f_2^2 = 9 T^2$ this means that $T\to 0$ and $R\to 1$ as the boundary is being approached.
Regularity of the solution at the boundary then requires the following behavior as $r \equiv 1-R  \to 0$, 
\bea
\label{eq:BCs} %(\ref{eq:BCs})
\kappa^{2} = \cO(r)
\hskip 0.8in  
\cG  = \cO (r^{3}) 
\hskip 0.8in 
\partial_{w}\cG  = \cO(r^{2}) 
\eea
The explicit expression for $R$ in terms of $\kappa^{2} \cG$ and $\p_w \cG$ in (\ref{eq:comp}) furthermore requires, 
\begin{align}
\label{eq:RBC} %(\ref{eq:RBC})
\frac{\kappa^{2}\mathcal{G}}{|\partial_{w}\mathcal{G}|^{2}} \to -\frac{2}{3}
\end{align}
Note that the boundary condition $\partial_{w}\mathcal{G} = 0$ on $\partial\Sigma$ is stronger than the corresponding condition for the $AdS_6$ case, where $(\partial_{w}+\partial_{\bar w})\mathcal{G}=0$ was sufficient \cite{D'Hoker:2016rdq}.

\subsection{Realizing the regularity conditions at the boundary \texorpdfstring{$\p \Sigma$}{dSigma}}
\label{sec:bc-sym}

The boundary conditions discussed in sec.~\ref{sec:bc-review} can be realized naturally  by imposing a conjugation condition on the holomorphic functions $\mathcal{A}_{\pm}$ on $\p \Sigma$. We shall take $\p \Sigma$ to consist of only a single connected boundary component, though the construction may be easily generalized to the case when $\p \Sigma$ has several components. We may map the boundary $\p \Sigma$ to the real line by a Schwarz-Christoffel transformation, which is piecewise conformal. Let $w,\bar w$ be local complex coordinates  in terms of which a boundary segment is given by $w=\bar w$. The conjugation condition is then given by, 
\begin{align}
\label{eq:conjug} %(\ref{eq:conjug})
\overline{\mathcal{A}_{\pm}(\bar{w})} = \mathcal{A}_{\mp}(w)
\end{align}
This condition readily implies $\kappa^{2} = 0$ on $\partial\Sigma$ and, noting that we have,
\begin{align}
\partial_{w}\cG(w,\bar w)  = (\overline{\cA_{+}(w)} - \mathcal{A}_{-}(w) )\partial_{w}\mathcal{A}_{+}(w) 
+ (\mathcal{A}_{+}(w) - \overline{\cA_{-}(w)} )\partial_{w}\mathcal{A}_{-}(w)
\end{align}
it also implies $\partial_{w}\mathcal{G} = 0$ on $\partial\Sigma$. Consequently, $\cG$ is constant along each boundary segment  and can be made to vanish on any one single  segment by fixing the integration constant implicit in the definitions of $\cB$ and $\cG$. The behaviors near the boundary in (\ref{eq:BCs}) are implied by the relations between $\mathcal{G}$, $\partial_{w}\mathcal{G}$, and $\kappa^{2}$ via differentiation, which in turn imply (\ref{eq:RBC}).

\sm

We conclude this section by drawing a comparison between the boundary conditions for the $AdS_2 \times S^6$ case studied here  and the boundary conditions for the $AdS_6 \times S^2$ case studied in \cite{DHoker:2017mds}.
The conjugation relation between the differentials resulting from (\ref{eq:conjug}) differs from the analogous condition for the differentials in the $AdS_6\times S^2$ solutions of \cite{DHoker:2017mds} only by a sign. 
More importantly, it was sufficient in \cite{DHoker:2017mds} to implement a conjugation condition on the differentials $\p_w\cA_\pm$ to ensure $(\partial_w+\partial_{\bar w})\cG\vert_{\partial\Sigma}=0$, whereas here we impose the conjugation relation on the functions $\cA_\pm$ themselves in order to implement the stronger condition $\partial_w\cG\vert_\Sigma=0$.

 Furthermore, the conjugation condition of (\ref{eq:conjug}) is incompatible with the presence of logarithmic branch cuts in $\cA_\pm$ starting at branch points on the boundary $\p \Sigma$ and with branch cuts along the boundary. Suppose that we have a branch point at $w=0$,
\begin{align}
 \cA_\pm (w) &=\cA_\pm^{(0)} (w) +\cA_\pm^{(1)}(w) \, \ln w 
 \hskip 1in
 \overline{\cA_\pm^{(i)}(\bar w)}=\cA_\mp^{(i)}(w)
\end{align}
where $\cA_\pm^{(0)}(w)$ and $\cA_\pm^{(1)}(w)$ are regular and single-valued in a neighborhood of $w=0$. 
Upon encircling $w=0$ counterclockwise, $\cA_\pm\rightarrow \cA_\pm+i\pi\cA_\pm^{(1)}$. 
This is compatible with (\ref{eq:conjug}) and the assumed conjugation properties of $\cA_\pm^{(i)}$ only if $\cA_\pm^{(1)}$ is zero as a function. Hence such branch cuts are ruled out, contrary to the case of $AdS_6 \times S^2$ where they were crucial ingredients in the construction of the global solutions.

%%%%%%%%%%%%%%%%%%%%%%%%%%%%%%%%%%%%%%%%%%%%%%%%%%
%%%%%%%%%%%%%%%%%%%%%%%%%%%%%%%%%%%%%%%%%%%%%%%%%%
\section{Towards string junction solutions}
\label{sec:string-ansatz}
%%%%%%%%%%%%%%%%%%%%%%%%%%%%%%%%%%%%%%%%%%%%%%%%%%
%%%%%%%%%%%%%%%%%%%%%%%%%%%%%%%%%%%%%%%%%%%%%%%%%%

In this section we determine the behavior needed for the functions $\cA_\pm$ to source the seven-form charges associated with  $(p,q)$-strings. We shall show that, in addition to reproducing the charges,  $\cA_\pm$ with this behavior correctly reproduces the metric, axion-dilaton, and two-form fields of the near-horizon limit of the classic $(p,q)$-string solution, provided we carry out a certain coordinate inversion to be explained below. Though we will be able to write down the functions $\cA_\pm$ producing $(p,q)$-string charges at multiple points on $\p \Sigma$, the question of whether these supergravity solutions are actually geodesically complete for some choices of the parameters remains unsettled.

\subsection{Realizing the charge and the \texorpdfstring{$S^7$}{S7} of the \texorpdfstring{$(p,q)$}{(p,q)}-string  solution}
\label{app:forming-S7}

To realize a $(p,q)$-string charge in an $AdS_2\times S^6$ supergravity solution, we begin by determining the behavior of the functions $\cA_\pm$ near a point $b \in \p \Sigma$ where a $(p,q)$-string charge resides. A first ingredient is that the supergravity fields should be regular in a neighborhood of the point $b$ with $b$ itself removed, and the seven-form should support $(p,q)$ charge.  A second ingredient is the fact that the classic $(p,q)$-string solution exhibits a round $S^7$ in the directions transverse to the string. Noting that the metric function $f_6^2$ vanishes on $\p \Sigma$, we conclude that the $S^7$ is realized by a fibration of $S^6$ over a curve in $\Sigma$ which begins and ends on $\p \Sigma$. The angular dependence required to realize this fibration smoothly will constrain  the functions $\cA_\pm$. 

\sm

Consider a point $b\in \p \Sigma$ and local complex coordinates $w, \bar w$ which vanish at this point. Regularity and single-valuedness of the supergravity fields $f_2^2, f_6^2, \rho^2,$ and $B$ near $w=0$ require $\cA_\pm$ to be single-valued in a neighborhood of $w=0$, just as was the case for $AdS_6 \times S^2$ solutions. The extra condition that the factor $d\cC$ in $F_{(3)}$ be residue-free at $w=0$ ensures the absence of  five-brane charges and excludes logarithmic branch cuts emanating from $w=0$. Thus, we shall assume that $\cA_\pm$ has a Laurent expansion in $w$ at $w=0$. While $\cA_\pm$ and $\cB+\bar \cB$  are single-valued near $w=0$, this set-up still allows the factor $d\cM$ of $F_{(7)}$  to have a non-zero residue and thus to carry a non-zero $(p,q)$-string charge.

\sm

Next, we determine the order of the pole in $\cA_\pm$ by requiring a smooth $S^6$ slicing of $S^7$. We shall assume that $\cA_\pm$ has a pole at $w=0$ of order at most $p-1$, 
\begin{align}
 \cA_\pm (w) &=\frac{\alpha_\pm}{w^{p-1}}+\frac{\beta_\pm}{w^{p-2}}+\frac{\gamma_\pm}{w^{p-3}} + \cdots
\end{align}
The coefficients are constrained by (\ref{eq:conjug}), so that $\bar \alpha _\pm = \alpha _\mp$ and likewise for $\b_{\pm},\g_{\pm}$, which forces the orders of the poles in $\cA_\pm$ to coincide with one another. Whether a smooth 7-cycle is formed around the pole at $w=0$ can be inferred from the ratio $f_6^2/\rho^2$. In terms of polar coordinates $w=r e^{i\theta}$ near the pole, the metric (\ref{eq:metric}) may be written as,
\begin{align}
 ds^2&=f_2^2ds^2_{AdS_2}+4\rho^2\left ( dr^2+r^2d\theta^2+\frac{f_6^2}{4\rho^2}ds^2_{S^6}\right )
\end{align}
A smooth cycle is formed if $f_6^2/\rho^2$ is positive for $\theta\in(0,\pi)$ and approaches zero quadratically as $\theta\rightarrow 0$ and $\theta\rightarrow\pi$.
For $p=2$ no smooth 7-cycle is formed. For $p \geq 3 $ we find,
\begin{align}
 \frac{f_6^2}{4\rho^2}&=-\frac{3\cG}{2\kappa^2}=
3 r^2 \,  \frac{ (2 p-3) \sin \theta -  \sin   (2p \theta -3 \theta) }{2 (p-1) (p-2)(2p-3) \sin \theta } + \cO(r^3)
\end{align}
For $p=3$ we find $f_6^2/4\rho^2\approx r^2\sin^2\!\theta$, giving rise to a round $S^7$ from $S^6$ and the polar part of the metric on $\Sigma$. For integer $p>3$, a smooth 7-cycle is formed which is not a round $S^7$.
We conclude that the poles in $\cA_\pm$ must be double, with $p=3$.

\subsection{Supergravity fields near a double pole in \texorpdfstring{$\cA_\pm$}{cA-pm}}
\label{sec:local-string-match}

To obtain the supergravity fields near a double pole in $\cA_\pm$, we need the Laurent expansions of these functions to order $w^3$, 
\begin{align}
\label{Adouble}
\mathcal{A}_{\pm} = \frac{\alpha_{\pm}}{w^{2}} + \frac{\beta_{\pm}}{w} + \gamma_{\pm} + \delta_{\pm} w + \epsilon_{\pm} w^{2} + \chi_{\pm} w^{3} + \mathcal{O}(w^{4})
\end{align}
along with the conjugation condition implied by (\ref{eq:conjug}) so that  $\bar{\alpha}_\pm = \alpha_\mp$, etc.  
The first regularity condition is that $\kappa^2 >0$ in the interior of $\Sigma$. The leading behavior of $\kappa^2$ is obtained from (\ref{Adouble}), 
\bea
\label{kap}
\kappa^{2}  = \frac{2 \zeta_{\alpha\beta} \, \Im w}{|w|^{6}} + \mathcal{O}(|w|^{-3}) 
\hskip 20mm
\zeta_{\alpha\beta} = 2i(\alpha_{+}\beta_{-} - \alpha_{-}\beta_{+})
\eea
The conjugation conditions imply that $\zeta _{\a\b}$ is real, and positivity of $\kappa^2$ in the upper half-plane requires $\zeta _{\a\b} >0$.
In addition, the function $\cG$, and hence $\cB + \bar \cB$, must be single-valued. Computing $\cB$ in terms of (\ref{Adouble}), we find,
\begin{align}
\mathcal{B} = \frac{i\zeta_{\alpha\beta}}{6 w^{3}} + \frac{i\zeta_{\alpha\gamma}}{2 w^{2}} + \frac{i(3\zeta_{\alpha\delta} + \zeta_{\beta\gamma})}{2w} - 2i(2\zeta_{\alpha\epsilon} + \zeta_{\beta\delta})\ln w + \mathcal{O}(|w|)
\end{align}
with $\zeta_{\alpha\gamma}$ etc.\ defined in analogy with $\zeta_{\alpha\beta}$. Single-valuedness of $\cB+ \bar \cB$ requires the purely imaginary coefficient of $\ln w$ to vanish, 
\begin{align}
2\zeta_{\alpha\epsilon} + \zeta_{\beta\delta} &= 0
\end{align}
The functions $\cG$ and $T$, in terms of which the metric functions $f_2^2, f_6^2, \rho^2$  are given by (\ref{f2f6rho}), take the following form near $w=0$, 
\begin{align}
\label{GT}
 \mathcal{G} &\approx - \frac{4 \zeta_{\alpha\beta}(\Im w)^3}{3|w|^{6}}
 & T^{2} & \approx 4\xi |w|^{2}(\Im w)^{2} 
 & -\xi & = \frac{2\zeta_{\alpha\chi}+\zeta_{\beta\epsilon}}{\zeta_{\alpha\beta}} + \frac{\zeta_{\alpha\delta}^{2}}{\zeta_{\alpha\beta}^{2}}
\end{align}
The condition $\zeta _{\a\b} >0$, which already guaranteed $\kappa ^2 >0$, is seen to also guarantee that $\cG<0$, as is indeed required by the regularity of the supergravity solution. In addition, we impose the requirement $\xi >0$ to render $T$ positive. 

\sm

With these conditions fulfilled, the behavior of the functions $f_2^2, f_6^2, \rho^2$, and $\cC$ near the pole is given as follows in terms of polar coordinates $w = r e^{i\theta}$ near $w=0$, 
\begin{align}
\label{near}
\rho^{2} & \approx \frac{\zeta_{\alpha\beta}^{1/2}\xi^{1/4}}{r^{5/2}} & f_{6}^{2} & \approx 4 r^{2}\sin^{2}\theta\,\rho^{2} & f_{2}^{2} & \approx \frac{\zeta_{\alpha\beta}^{1/2}}{9\xi^{3/4}r^{9/2}}
& \mathcal{C} & \approx \frac{-2i\alpha_{+}}{9\xi r^{6}}
\end{align}
The complex axion-dilaton field $\tau$, for $\alpha_{+} \neq \alpha_{-}= \bar \alpha _+$, is given by, 
\begin{align}
\Re(\tau)  & \approx \frac{\Re (\alpha _+) }{\Im (\alpha_+)} & 
\textrm{Im}(\tau) &\approx \frac{\xi^{1/2}\zeta_{\alpha\beta}}{4 \, \Im (\alpha _+)^2 } \, r^{3}
\end{align}
Finally, the presence of string charge at the pole may be verified by examining the potential $\cM$ for the seven-form field strength $F_{(7)}$. By inspection of (\ref{eq:cC-cM}) we see that all terms in $\cM$ are single-valued by construction, except for the contributions from the locally holomorphic functions $\cW_\pm$, whose behavior near $w=0$ is given as follows, 
\begin{align}
\mathcal{W}_{\pm} (w) = \mathcal{W}_{\pm}^{s} (w) - \frac{3}{2}i \Big (  (3\zeta_{\alpha\chi} + \zeta_{\beta\epsilon})\beta_{\pm} + \zeta_{\alpha\delta}\delta_{\pm} - \zeta_{\alpha\beta}\chi_{\pm} \Big ) \, \ln w
\end{align}
where $\cW_\pm^s$ denotes the single-valued part.  Therefore, as $w$ encircles the pole at $w=0$ counterclockwise in $\Sigma$ by a $180^\circ$ degree arc, the potential $\cM$ shifts as follows, 
\begin{align}
\mathcal{M} \to \mathcal{M} + 240\pi \Big (  (3\zeta_{\alpha\chi} + \zeta_{\beta\epsilon})\beta_+ + \zeta_{\alpha\delta}\delta_+ - \zeta_{\alpha\beta}\chi_+ \Big ) 
\end{align}
The shift in $\cM$ gives the integral of the seven-form field strength over the $S^7$, producing 
a formula for the $p,q$ string charges in terms of the coefficients of the Laurent series,
\bea
p+iq = \int _{S^7} F_{(7)} = 80 \pi^5 \Big (  (3\zeta_{\alpha\chi} + \zeta_{\beta\epsilon})\beta_+ + \zeta_{\alpha\delta}\delta_+ - \zeta_{\alpha\beta}\chi_+ \Big ) 
\eea
where we have used ${\rm vol} (S^7) =\pi^4/3$. 
Note that the dependence of the string charges $p,q$ on the coefficients of the Laurent series is trilinear, in contrast with the $AdS_6 \times S^2$ case where the  five-brane charges had linear dependence.

\subsection{Satisfying the regularity conditions near a double pole}

The various positivity and regularity conditions derived in the preceding subsection may be satisfied simultaneously. To see this, we use the $SU(1,1)$ symmetry of supergravity to rotate $\alpha _\pm$ to be real, and furthermore scale it to 1 without loss of generality. The conditions then reduce to the following relations, 
\begin{align}
\zeta _{\a\b} &= 4 \, \Im (\beta_+) >0\, , &  \Im (\epsilon_+) &= -{1 \over 8}  \zeta _{\beta \delta}\, , 
&  \xi &=  - {  (\Im \delta_+)^2 \over (\Im \beta_+)^2} - {8 \, \Im \chi_+ + \zeta _{\beta \epsilon}  \over 4 \, \Im \beta_+} >0
\end{align}
For given $\Re(\beta_+)$, $\Im (\beta _+)>0$, $\delta _+$, and $\ep_+$, we may always choose $\Im (-\chi_+)$ large enough to satisfy the remaining condition $\xi >0$. The expression for the charge $p$ is unenlightening, but the charge $q$ takes the simple form  $q = -320 \pi^5 \, \xi \, (\Im \beta )^2$ and must be negative.

\sm

We conclude this subsection with a remark on  a no-go result  derived in \cite{Corbino:2017tfl}. 
Assuming certain regularity conditions on $\kappa^2$ and $\cG$, it was argued  that $\kappa^2>0$ and $\cG<0$ cannot both be realized for compact $\Sigma$ with boundary. This argument was based on  an integral representation for  $\cG$, obtained by solving the differential  relation $\kappa^2=-\partial_w\partial_{\bar w}\cG$, 
\begin{align}
\label{eq:5_5_1b}% (\ref{eq:5_5_1b})
\mathcal{G}(w) &= H(w) + \frac{1}{\pi}\int_{\Sigma}d^{2}z~G(w,z)\kappa^{2}(z)
\end{align}
with harmonic $H$ and $G$ the Green's function on $\Sigma$. The functions $\kappa ^2$ and $\cG$ obtained here circumvent this no-go result, because they are too singular at the pole to allow for the  integral representation (\ref{eq:5_5_1b}).  Indeed,  the singularity in $\kappa^2 $ is not integrable against the Green function, as may be seen from the form of $\kappa ^2$ and $\cG$ given in (\ref{kap}) and (\ref{GT}).  This shows that the assumptions entering the argument of sec.~7.3.2 of \cite{Corbino:2017tfl} do not hold here.

\subsection{Matching with the classic \texorpdfstring{$(p,q)$}{(p,q)}-string solutions}
\label{app:string-sol-limit}

The classic $(p,q)$-string solutions of Type IIB supergravity constructed in \cite{Schwarz:1995dk} are labeled by a pair of integers $(q_{1},q_{2})$ which characterize the charges.  The metric, two-form, and axion-dilaton $\tau = \chi + i e^{-\phi}$ are given by,
\begin{align}
\label{eq:SchwarzPQ} %(\ref{eq:SchwarzPQ})
ds^{2} & = A_{q}^{-3/4}ds_{\mathds{R}^{1,1}}^{2} + A_{q}^{1/4}\left( dy^{2} + y^{2}d{s}_{S^{7}}^{2} \right) 
&
\tau & = \frac{q_{1}\chi_{0} - q_{2}|\tau_{0}|^{2} + i q_{1}e^{-\phi_{0}}A_{q}^{1/2}}{q_{1} - q_{2}\chi_{0} + i q_{2}e^{-\phi_{0}}A_{q}^{1/2}}
\nonumber \\
B_{01}^{(i)} & = \left( \mathcal{M}_{0}^{-1} \right)_{ij} q_{j}\Delta_{q}^{-1/2} A_{q}^{-1} 
& A_{q} &= 1 + \frac{\alpha_{q}}{y^{6}}
\end{align}
The asymptotic values of the axion-dilaton are given by $\tau_0 = \chi_0 + i e^{-\phi_0}$, and we have, 
\begin{align}
\label{eq:SchwarzAq} %(\ref{eq:SchwarzAq})
\alpha_{q} & = \Delta_{q}^{1/2} Q 
& \Delta_{q} & =  \begin{pmatrix}   q_{1}\\  q_{2} \end{pmatrix}^t 
\mathcal{M}_{0}^{-1}
\begin{pmatrix}   q_{1}\\  q_{2} \end{pmatrix}
&
\mathcal{M} & = e^{\phi}
\begin{pmatrix}   |\tau|^{2}       & \chi \\   \chi  & 1 \end{pmatrix}
\end{align}
As $y \to \infty$ we recover flat spacetime $\RR^{1,9}$.  The near-horizon limit corresponds to $y^6 \ll \alpha_q$, so that the first term in $A_q$ may be neglected in this limit and we have simply $ A_q (y)\to \alpha_{q}/y^6$. The supergravity fields take the following form, 
\begin{align}
ds^{2} &= {y^{{9 \over 2}} \over \a_q^{3/4}} ds_{\mathds{R}^{1,1}}^{2} 
+  {\a_q^{1/4} \over y^{{3 \over 2}}} ( dy^{2}  + y^2 ds_{S^{7}}^{2}) 
&
\tau & = \frac{q_{1}\chi_{0} - q_{2}|\tau_{0}|^{2} + i q_{1}e^{-\phi_{0}}\sqrt{\alpha_q}/y^3}{q_{1} - q_{2}\chi_{0} + i q_{2}e^{-\phi_{0}}\sqrt{\alpha_q}/y^3}
\no\\
B_{01}^{(i)} &= \left( \mathcal{M}_{0}^{-1} \right)_{ij} q_{j}\Delta_{q}^{-1/2}\frac{y^6}{\a_q}
\end{align}
In this limit, $ y \to 0$ corresponds to the location of the string, but this is a strong coupling limit since the dilaton blows up there. The limit $y \to \infty$ corresponds to the other end of the throat which is also a strong coupling limit.
Clearly, identifying the coordinate $r$ of (\ref{near}) with $y$ does not lead to a match between the supergravity fields of the $AdS_2 \times S^6$ solutions and the supergravity fields of the classic $(p,q)$-string solution to Type IIB. However, if we perform a coordinate inversion on $y$ in the string solution by setting,
\bea
\label{inversion}
y = L /r
\eea
then the supergravity fields of the string solution  in terms of $r$ are given by, 
\begin{align}
ds^{2} &= \frac{L^{{9 \over 2}} ds_{\mathds{R}^{1,1}}^{2}}{ \a_q^{3/4}r^{9/2}} + \frac{L^\half \a_q^{1/4}}{r^{5/2}}(dr^{2} + r^{2}ds_{S^{7}}^{2}) 
&
\tau & = \frac{q_{1}\chi_{0} - q_{2}|\tau_{0}|^{2} + i q_{1}e^{-\phi_{0}}\sqrt{\alpha_q}r^3/L^3}{q_{1} - q_{2}\chi_{0} + i q_{2}e^{-\phi_{0}}\sqrt{\alpha_q}r^3/L^3}
\no\\
B_{01}^{(i)} &= \left( \mathcal{M}_{0}^{-1} \right)_{ij} q_{j}\Delta_{q}^{-1/2}\frac{L^6}{\a_q r^{6}}
\end{align}
which perfectly match with the $AdS_2 \times S^6$ solution provided we identify the parameters,
\bea
L^3 = { \zeta _{\alpha \beta} \over 3} 
\hskip 1in 
\alpha _q = \xi \, ( 3 \zeta _{\a\b} ^2)^{{2 \over 3}}
\eea
and a corresponding identification for the flux field.
Note that the worldvolume for the $AdS_2 \times S^6$ solution is $AdS_{2}$, whereas for the classic string solution it is $\mathds{R}^{1,1}$. The inversion in the identification (\ref{inversion}) may play 
a role in the physical interpretation of potential global solutions.

\subsection{Multiple \texorpdfstring{$(p,q)$}{(p,q)} charge solutions on the upper half-plane}

In the previous subsection, we have shown that a double pole in the functions $\cA_\pm$ on the boundary $\p \Sigma$ produces supergravity fields which may be identified locally, i.e. in a finite neighborhood of the pole, with the supergravity fields of the classic $(p,q)$-string solution. Here we shall extend this construction to the case of multiple double poles which are all located on the boundary $\p \Sigma$. For simplicity, we shall consider the case where $\Sigma$ has the topology of the upper half-plane, for which the boundary is the real line. Hence we shall consider functions $\cA_\pm$ with $N$ double poles, located at points $p_\ell \in \RR$ for $\ell =1, \cdots, N$. 

\sm

To make further progress, we shall assume that $\cA_\pm$ are rational functions  of $w$ and that $w=\infty$ is a regular point (which may always be achieved by conformal mapping). The functions may therefore be decomposed into partial fractions in $w$ as follows, 
\bea
\cA_\pm = \cA_\pm ^{(0)} + \sum _{\ell=1}^N \left ( { Y_\pm ^\ell \over (w-p_\ell)^2} + { Z_\pm ^\ell \over w-p_\ell} \right )
\eea
where $Y^\ell _- = \bar Y^\ell _+$, $Z^\ell _- = \bar Z^\ell _+$, and $\cA_\pm ^{(0)} $ are complex parameters which are independent of~$w$. This Ansatz implements the reflection condition (\ref{eq:conjug}), as a result of which $\kappa ^2 $ and $\cG$ vanish on $\p \Sigma = \RR$. It remains to enforce the positivity requirement $\kappa^2>0$ everywhere  in the interior of the upper half-plane, which in particular requires that $\p \cA_-$ has no zeros in the upper half-plane. We also need the condition that the function $\cB + \bar \cB$ be single-valued.  
\sm

An alternative formulation starts from the differentials $\p \cA_\pm$, which have a triple pole at each $w=p_\ell$. We may easily enforce the conditions that the zeros of $\p \cA_+$ and $\p \cA_-$ all be located in the upper and lower half-planes, respectively, by the following parametrization (analogous to the parametrization used for the $AdS_6$ case in \cite{DHoker:2017mds}), 
\bea
\label{eq:disc1} %(\ref{eq:disc1})
\partial_{w}\mathcal{A}_\pm  = P_\pm (w) \prod_{\ell=1}^{N}\frac{1}{(w-p_{\ell})^{3}} 
\hskip 0.4in 
P_+ (w)  = \prod_{n=1}^{3N-2}(w - s_{n})  \hskip 0.3in P_- (w)  = \prod_{n=1}^{3N-2}(w - \bar s_{n})
\quad
\eea
with $\Im (s_n) >0$. In order to integrate to single-valued functions $\cA_\pm$ and $\cB+ \bar \cB$, the differentials $\p \cA_\pm$ must have vanishing residues at $p_\ell$, while the imaginary part of the residue of the differential $\p_w \cB$  must also vanish, 
\bea
\label{eq:disc3} %(\ref{eq:disc3})
\mathrm{Res}(\partial_{w} \mathcal{A}_{\pm}) \Big |_{w=p_{\ell}} = 0 \hskip 1in 
 \mathrm{Res}(\partial_{w}\mathcal{B}) \Big |_{w=p_{\ell}}  \in \RR
\eea
The counting of parameters shows that, for a given arrangement of poles $p_\ell$, there are $3N-2$ complex zeros, subject to $3N-3$ real residue conditions. Thus parameter counting allows for the existence of large families of solutions.  While it is clear that the positivity and regularity conditions are satisfied in the neighborhood of each pole, and that the supergravity fields match onto a classical $(p,q)$-string solution in the near-horizon limit, it is unclear how to ensure regularity throughout the upper half-plane. The solutions found numerically thus far have all been geodesically incomplete, and this includes the cases with one and two charges. The situation will be discussed explicitly for the case of three charges in the next subsection.

\subsection{Three charges}

In this final subsection, we analyze the case of three double poles in $\cA_\pm$, and thus three $(p,q)$-string charges.
In order to conveniently exploit as much symmetry of the configuration as possible, we shall work on the unit disc with complex coordinates $z, \bar z$ rather than on the upper half-plane with complex coordinates $w, \bar w$. The conjugation condition (\ref{eq:conjug})  on the disc becomes $\overline{\cA_\pm (1/\bar z) } = \cA_\mp(z)$, and we may exploit $SU(1,1)$ symmetry of the unit disc to map the positions of the poles to $1, \ep, \ep^2$ where $\ep$ is a non-trivial cube root of unity. The differentials and polynomials of (\ref{eq:disc1}) then take the form,
\bea
\p_z \cA_\pm (z) = { P_\pm (z) \over (z^3-1)^3} 
\hskip 0.6in 
P_+(z) = \sum _{k=0}^7 c_k z^k 
\hskip 0.4in 
P_-(z) = \sum _{k=0}^7 \bar c_{7-k} z^k
\eea
The vanishing of the residues of $\p_z \cA_\pm$ at the poles gives two complex linearly independent relations between the coefficients $c_k$, 
\bea
\label{eq:resA}
c_3 & = & 5 c_0 + 2 c_6
\no \\
c_4 & = & 5 c_7 + 2 c_1
\eea
while the vanishing of the imaginary part of the residues of $\p_z \cB$ gives two  independent real  relations, which may be combined into one complex relation between the coefficients $c_k$,
\bea
0 & = & 27 |c_7|^2 + 9 |c_6|^2 - 2|c_5|^2 + 2 |c_2|^2 - 9 |c_1|^2 -27 |c_0|^2 
\no \\ &&
-18 \bar c_7 c_6 +21 \bar c_2 c_7 -9 \bar c_1 c_7-36 \bar c_7 c_1 -3 \bar c_6 c_2
\no \\ &&
 + 9 \bar c_0 c_6 + 36 \bar c_6 c_0 + 3 \bar c_5 c_1 -21 \bar c_0 c_5 +18 \bar c_1 c_0
\eea 
where we have eliminated $c_3, c_4$ using (\ref{eq:resA}). Global regularity and geodesic completeness requires furthermore that we have $\kappa ^2 >0, \cG<0$, and $T$ real. The condition $\kappa^2>0$ in the interior of the disc requires that all the zeros of $P_+(z)$ be in the interior of the disc, which implies that all the zeros of $P_-(z)$ will be outside the disc. We have not been able to solve this condition in any general form, nor numerically for any particular choice of parameters~$c_k$. However, we have also not been able to show convincingly that no solutions exist. The cases with 4 poles have also been explored, but the complexity of the conditions required is then even more involved.
In the absence of these results, we are left only with solutions with $(p,q)$-string charges which are not geodesically complete.

%%%%%%%%%%%%%%%%%%%%%%%%%%%%%%%%%%%%%%%%%%%%%%%%%%
%%%%%%%%%%%%%%%%%%%%%%%%%%%%%%%%%%%%%%%%%%%%%%%%%%
\section{Non-compact globally regular solutions}
\label{sec:global-sol}
%%%%%%%%%%%%%%%%%%%%%%%%%%%%%%%%%%%%%%%%%%%%%%%%%%
%%%%%%%%%%%%%%%%%%%%%%%%%%%%%%%%%%%%%%%%%%%%%%%%%%

In this section, we shall study solutions independently from a string junction interpretation and construct families of globally regular and geodesically complete solutions with asymptotic regions where spacetime decompactifies.

\subsection{Poles in the interior of \texorpdfstring{$\Sigma$}{Sigma}}
\label{sec:poles-in-Sigma}

In the case of $AdS_6$ solutions, poles in the interior of $\Sigma$ were not compatible with the regularity conditions \cite{DHoker:2017mds}.  In contrast, due to $\kappa^2$ and $\cG$ having opposite signs, poles in the interior  of $\Sigma$ can be realized for $AdS_2$ solutions. Positivity of $\kappa^2$ requires that at any singular point the divergence in $\partial_w\cA_-$ be at least as strong as the one in $\partial_w\cA_+$. We start with the case where $\partial_w\cA_+$ is subleading with respect to $\partial_w\cA_-$ and we shall generalize afterwards.

\sm

With $\cA_+$ subleading to $\cA_-$, the leading behavior of the composite quantities is given by the behavior of $\cA_-$ as follows,
\begin{align}
 \kappa^2&\approx |\partial_w\cA_-|^2
 &
 \cG&\approx -|\cA_-|^2
 &
 \partial_w\cG&\approx -\bar \cA_-\partial_w\cA_-
\end{align}
As the pole is approached this yields $T^2\approx \frac{1}{3}$. The metric functions, axion-dilaton scalar $B$, 
and the functions $\cC$ and $\cM$ parametrizing the two- and six-form potentials become,
\begin{align}
 f_2^2&\approx \sqrt[4]{\frac{4}{27}}\ |\cA_-|
 &
 f_6^2&\approx \sqrt[4]{12}\:|\cA_-|
 &
 \rho^2&\approx\frac{1}{\sqrt[4]{4\cdot 27}}\frac{|\partial_w\cA_-|^2}{|\cA_-|}
\no\\
 B&\approx-\left(2-\sqrt{3}\right)\frac{\bar\cA_-}{\cA_-}
 &
 \cC&\approx \frac{4i}{3}\bar\cA_-
 &
 \cM&\approx 8|\cA_-|^2\bar\cA_-
\end{align}
In particular, we have $|B|<1$, such that $\Im(\tau)$ is non-zero and finite. Assuming that $\cA_-$ has a pole of order $n$ with a complex coordinate $z$ centered on the pole, we have, 
\begin{align}\label{eq:cAm-pole}
 \cA_-&\approx \frac{a}{z^n} & \partial_w\cA_-&\approx -\frac{na}{z^{n+1}}
\end{align}
Since $\Im(\tau)$ is finite and non-zero, the metric in string-frame is related to the Einstein-frame metric by a finite rescaling. 
The Einstein-frame metric near the pole becomes,
\begin{align}
 ds^2&\approx  \frac{\sqrt{2}|a|}{3^{3/4}|z|^n}\left ( ds^2_{AdS_2}+3ds^2_{S^6}+2n^2\left|\frac{dz}{z}\right|^2\right )
\end{align}
The proper distance between the point $z=0$ and any other point on $\Sigma$ is infinite. 
This suggests a change of coordinates on $\Sigma$ to $u=1/z$,\footnote{This transformation will be part of the $SL(2,\CC)$ and $SL(2,\RR)$ automorphisms of the sphere and the upper half plane, respectively, in the examples to be discussed below.} such that the near-pole region corresponds to $|u|\gg 1$. The metric becomes, 
\begin{align}\label{eq:pole-metric}
 ds^2&\approx  \frac{\sqrt{2}|a|}{3^{3/4}}\left ( |u|^n ds^2_{AdS_2}+3|u|^n ds^2_{S^6}+2n^2\left|u^{n/2-1}du\right|^2\right )
\end{align}
At the pole, the radii of $AdS_2$ and $S^6$ diverge.
The geometry decompactifies; it is regular and asymptotically conical. For $n=1$ it approaches the asymptotic region of a cone with deficit angle $\pi$, for $n=2$ there is no deficit angle, and for $n>2$ it has an excess angle. The singularity at the apex of the cone $u=0$ is not a problem since this form of the metric is only valid in the regime $|u| \gg 1$. 

The two- and six-form RR potentials are given by,
\begin{align}
 C_{(2)}&\approx\frac{4i}{3}\bar a \left(\frac{\bar u}{u}\right)^{n/2} |u|^{n} \vol_{AdS_2}
 &
 C_{(6)}&\approx 8|a|^2\bar a \left(\frac{\bar u}{u}\right)^{n/2} |u|^{3n}\vol_{S^6}
\end{align}
The $|u|^{n}$ and $|u|^{3n}$ divergences in $\cC$ and $\cM$, respectively, combine with the volume forms on unit-radius $AdS_2$ and $S^6$ to the volume forms that are naturally associated with the $AdS_2$ and $S^6$ factors of radius proportional to $|u|^{n/2}$ in the metric (\ref{eq:pole-metric}). The remaining coefficient functions in $C_{(2)}$ and $C_{(6)}$ are finite and regular.
With $u=re^{i\theta}$, the axion-dilaton scalar asymptotes to,
\begin{align}
 \tau&\approx\frac{\sin(2n\theta)+i\sqrt{3}}{2-\cos(2n\theta)} 
\end{align}
showing again that the dilaton is finite at the pole, with non-trivial dependence on the angular coordinate. The entire solution is thus regular at the pole.

\sm

The asymptotic metric has an additional $U(1)$ isometry, acting as phase transformations on $z$ and $u$. This symmetry is broken by the remaining fields $\tau$, $C_{(2)}$, and $C_{(6)}$ to a $\ZZ_n$ symmetry acting as $z\rightarrow e^{2\pi i/n}z$, or $u\rightarrow e^{-2\pi i/n}u$, which leaves $\cA_-$ in (\ref{eq:cAm-pole}) and the entire solution invariant. Whether this symmetry in the asymptotic region extends to a symmetry of the full solution depends on the precise form of $\cA_\pm$.

\sm

Finally, we note that this regularity analysis generalizes to the case where $\cA_+$ and $\cA_-$ both have poles, of the form
\begin{align}
 \cA_+&\approx \frac{b}{z^n} & \cA_-&\approx \frac{a}{z^n}
\end{align}
with $|a|>|b|$. An $SU(1,1)$ transformation (\ref{eq:SU11}) with $v=bu/a$ and arbitrary $u$ such that $|u|^2=|a|^2/(|a|^2-|b|^2)$ then sets $b=0$, and the analysis reduces to the one presented above. The Einstein-frame metric is invariant under $SU(1,1)$, so the previous discussion applies directly. Axion-dilaton scalar and RR-potentials transform but remain regular, and the action of the $\ZZ_n$ symmetry is unchanged.

\subsection{Solutions for \texorpdfstring{$\Sigma$}{Sigma} without boundary}

Having found local regularity in the vicinity of an interior pole, we now try to embed this in a globally regular solution. We consider a compact Riemann surface $\Sigma$ of genus $g$ without boundary, and investigate the positivity and regularity conditions on $\cA_\pm$ for such a surface. Since there is no boundary, there is no need for a conjugation relation between $\cA_\pm$. The number of zeros $M$ and the number of poles $N$ of the meromorphic 1-form differentials $ \p_w \cA_\pm$ on $\Sigma$ are related to the genus $g$ by the Riemann-Roch theorem, 
\begin{align}
 M &=N+2g-2
\end{align}
For surfaces with $g> 1$ the differentials $\p_w \cA_\pm$ necessarily have zeros in $\Sigma$. To keep  $\kappa^2$ non-negative, any  zero of $\partial_w \cA_-$ must also be a zero of $\partial_w\cA_+$. However, this means that $\kappa ^2$ vanishes at these zeros,  generally leading to solutions with conical singularities. These zeros can be avoided for genus zero surfaces without boundary, as we now discuss, as well as for solutions with boundary, to be discussed in the next subsection. 
 
 \sm

We take $\Sigma$ to be the sphere and assume that infinity is a regular point of $\cA_\pm$. Single-valued differentials $\partial_w\cA_\pm$ without zeros can only have one double pole. Implementing the condition $\kappa ^2 >0$ on $\cA_\pm$ and its differentials, we find, 
\begin{align}\label{eq:Apm-Sphere}
 \cA_\pm&=a_\pm -\frac{b_\pm}{w-p} &  \partial_w\cA_\pm&= \frac{b_\pm}{(w-p)^2} & |b_+|&<|b_-|
\end{align}
An $SU(1,1)\otimes \CC$ transformation (\ref{eq:SU11}) allows us to set $b_+=a_+=0$, so that we have, 
\begin{align}
 \kappa^2&=|\partial_w\cA_-|^2
 &
 \cG&=\cG_0-\left|\cA_-\right|^2
 &
 T^2&=1- \frac{2 \, |\cA_-|^2 }{3 \left (|\cA_-|^2 - \cG_0 \right )}
\end{align}
with integration constant $\cG_0$. Manifestly, $\kappa^2$ is positive  and $\cG$ is negative throughout $\Sigma$ provided $\cG_0<0$. With this choice we have $1/3<T^2<1$, such that the regularity condition for $R$ is satisfied as well.
Since $\kappa^2$, $\cG$, and $T^2$ are invariant under the $SU(1,1)\otimes \CC$ transformations (\ref{eq:SU11}), this implies that the entire class of solutions (\ref{eq:Apm-Sphere}) is regular for appropriate choices of $\cG_0$.
With the analysis of sec.~\ref{sec:poles-in-Sigma} for the pole, we have thus realized globally regular $AdS_2\times S^6$ solutions. The geometry decompactifies at the pole and approaches the asymptotic region of a cone with deficit angle $\pi$. The axion-dilaton scalar $B$ and the functions $\cC$ and $\cM$  are single-valued for all solutions (\ref{eq:Apm-Sphere}), and therefore carry no charges.

\sm

The Ansatz (\ref{eq:Apm-Sphere}) has one complex parameter in each of $a_\pm$, $b_\pm$, $p$,  and an additional real parameter  in $\cG_0$, making for $11$ real parameters. Subtracting $3$ complex parameters for the $SL(2,\CC)$  automorphisms on the sphere  leaves a total of 5 real parameters. The $SU(1,1)\otimes \CC$ duality transformations (\ref{eq:SU11}) map the class of functions (\ref{eq:Apm-Sphere}) into itself.

\subsection{Solutions  for \texorpdfstring{$\Sigma$}{Sigma} with boundary}\label{sec:disc-interior-pole}

The presence of a boundary for $\Sigma$ gives greater flexibility for the distribution of zeros and poles of $\cA_\pm$ compatible with the positivity and regularity conditions. We take $\Sigma$ to be the upper half-plane with the real line as its boundary. Assuming that the differentials $\p_w \cA_\pm(w)$ are rational functions 
of $w$, the condition $\kappa ^2 >0$ may be solved by an electro-statics problem  \cite{DHoker:2017mds,DHoker:2017zwj} and the solution is given by, 
\begin{align}
\label{eq:pot}
\frac{\partial_w\cA_+}{\partial_w\cA_-}=\lambda_0^2\prod_{n}\frac{w-s_n}{w-\bar s_n}
\end{align}
where $s_n$ are points in the upper half-plane with $\Im (s_n) >0$ and $\lambda _0$ is a constant. In the case of $AdS_6$ solutions, no poles in the interior of $\Sigma$ were allowed, which forced us to assign all points $s_n$ to be zeros of $\p_w \cA_+$ and by conjugation all points $\bar s_n$ to be zeros of $\p_w \cA_-$. However, common poles on the real line were allowed. In contrast, in the case of $AdS_2$ solutions we can allow for poles in the interior of $\Sigma$. To have single-valued $\cA_\pm$ functions, the poles must be of order at least two. Thus, we may distribute the points $s_n$ amongst the zeros of $\p_w \cA_+$ and the poles of $\p_w\cA_-$, and allow for additional zeros and poles on the real axis common to $\cA_\pm$. The general form of the differentials is then as follows, 
\begin{align}
 \partial_w\cA_+&= \prod_{k=1}^{N} \frac{1}{(w-\bar t_k)^{\nu_k}} \times \prod_{m=1}^{N_u} (w-u_m)
 \hskip 1in N_u =  \sum _{k=1}^{N} \nu_k -2
 \no\\
 \partial_w\cA_-&= \prod_{k=1}^{N} \frac{1}{(w-t_k)^{\nu_k}} \times \prod_{m=1}^{N_u} (w-\bar u_m)
 \hskip 1in \Im (t_k), \Im (u_m) \geq 0
\end{align}
The regularity conditions required to have a single-valued $\cA_\pm$ and $\cG$  are given as follows,
\begin{align}
\label{eq:Apm-res}
\mathrm{Res}(\partial_w\cA_-) \Big | _{w=t_k} &=0 & 
 \mathrm{Res}(\partial_w\cB) \Big | _{w=t_k} &\in\RR 
\end{align}
for $k=1,\ldots,N$. In addition one has to ensure that $\cG<0$ throughout the interior of $\Sigma$ with $\cG=0$ on $\partial\Sigma$ and that $0<T^2<1$.\footnote{For the $AdS_6$ case, regularity of $\kappa^2$ together with the boundary condition for $\cG$ automatically implied the full set of regularity conditions. This is not the case for $AdS_2$.} This construction introduces a large number of parameters and we show in the following subsections that regular solutions exist.

\sm

If the conditions (\ref{eq:Apm-res}) are satisfied, the axion-dilaton scalar and two-form potential do not have branch points at $w=t_k$ since $\cA_\pm$ are single-valued. In the six-form potential the terms involving $\cW_\pm$ in $\cM$ are potentially non-single-valued. However, due to the conjugation condition relating $\cA_\pm$, the residues of $\partial_w\cW_+$ at $w=t_k$ and of $\partial_{\bar w}\bar\cW_-$ at $\bar w=\bar t_k$ are related. As a result, even though individual terms in $\cM$ may have logarithmic branch cuts, the total monodromy of $\cM$ vanishes. There are thus no apparent brane charges at the poles $t_k$.

\subsection{Minimal solutions on the disc}

The minimal non-trivial case is $N=1$ with $\nu_1=2$,
\begin{align}
\label{minimaldA}
 \partial_w\cA_+&=\frac{\bar\sigma}{(w-\bar t_1)^2} & \partial_w\cA_-&=\frac{\sigma}{(w-t_1)^2}
\end{align}
The functions $\cA_\pm$ are given by
\begin{align}
\label{minA}
 \cA_+&=\cA_+^0-\frac{\bar\sigma}{w-\bar t_1} & 
 \cA_-&=\cA_-^0-\frac{\sigma}{w-t_1}
\end{align}
It will be convenient to map this solution to the disc, $\Sigma=\lbrace z\in\CC \,\big\vert\, |z|^2\leq 1\rbrace$. 
With redefined constants $\varsigma=-\sigma/(t_1-\bar t_1)$ and $\tilde\cA_+^0=\cA_+^0-\bar\varsigma$, $\tilde\cA_-^0=\cA_-^0-\varsigma$, the functions $\cA_\pm$ are then,
\begin{align}
\label{eq:cA-ex}
 \cA_+&=\tilde\cA_+^0+\bar\varsigma z & \cA_-&=\tilde \cA_-^0+\frac{\varsigma}{z}
 &
 z&=\frac{w-t_1}{w-\bar t_1}
\end{align}
They satisfy $\overline{\cA_\mp(1/\bar z)}=\cA_\pm$, which is the analog of the conjugation condition (\ref{eq:conjug}) on the disc.
For $\kappa^2$ and $\cG$ we find, with a suitable choice of the integration constant in $\cB$ to ensure $\cG\vert_{\partial\Sigma}=0$ and using that $\tilde\cA_+^0$ and $\tilde\cA_-^0$ are related by conjugation, 
\begin{align}
 \kappa^2&=|\varsigma|^2\left(\frac{1}{|z|^4}-1\right)
 &
 \cG&=|\varsigma|^2\left(|z|^2-\frac{1}{|z|^2}-2\ln|z|^2\right)
\end{align}
Manifestly, the positivity and regularity conditions $\kappa^2>0$ and $\cG<0$ are obeyed in the interior of $\Sigma$, and both functions vanish on the boundary. Moreover, we have,
\begin{align}
 \frac{\kappa^2\cG}{|\partial_z\cG|^2}&=\frac{1+|z|^2}{(1-|z|^2)^3}\left(|z|^4-1-2|z|^2\ln|z|^2\right)
\end{align}
This is a smooth function which monotonically increases from $-1$ at $|z|=0$ to $-2/3$ at $|z|=1$, as required by the boundary conditions. Therefore, these functions $\cA_\pm$ yield a solution which is regular everywhere. At the pole in $\cA_-$, we have that $\cA_+$ is finite. By the arguments of sec.~\ref{sec:poles-in-Sigma}, the geometry decompactifies at this point and approaches the asymptotic region of a cone with deficit angle~$\pi$.

\sm

Taking into account the conjugation relation between $\cA_+$ and $\cA_-$, the ansatz in (\ref{minA}) has 6 real parameters. The integration constant in $\cG$ is fixed by the boundary condition. Subtracting 3 degrees of freedom for redundancies due to the $SL(2,\RR)$ automorphisms of the upper half-plane leaves 3 free real parameters. We note that the $SU(1,1)\times\CC$ duality transformations (\ref{eq:SU11}) do not map the class of functions in (\ref{minA}) into itself; the shifts do, but $SU(1,1)$ transformations with $v\neq 0$ do not. The functions in (\ref{minA}) are therefore representatitves of entire $SU(1,1)$ orbits of non-trivial solutions.

\begin{figure}
\centering
\includegraphics[height=0.315\linewidth]{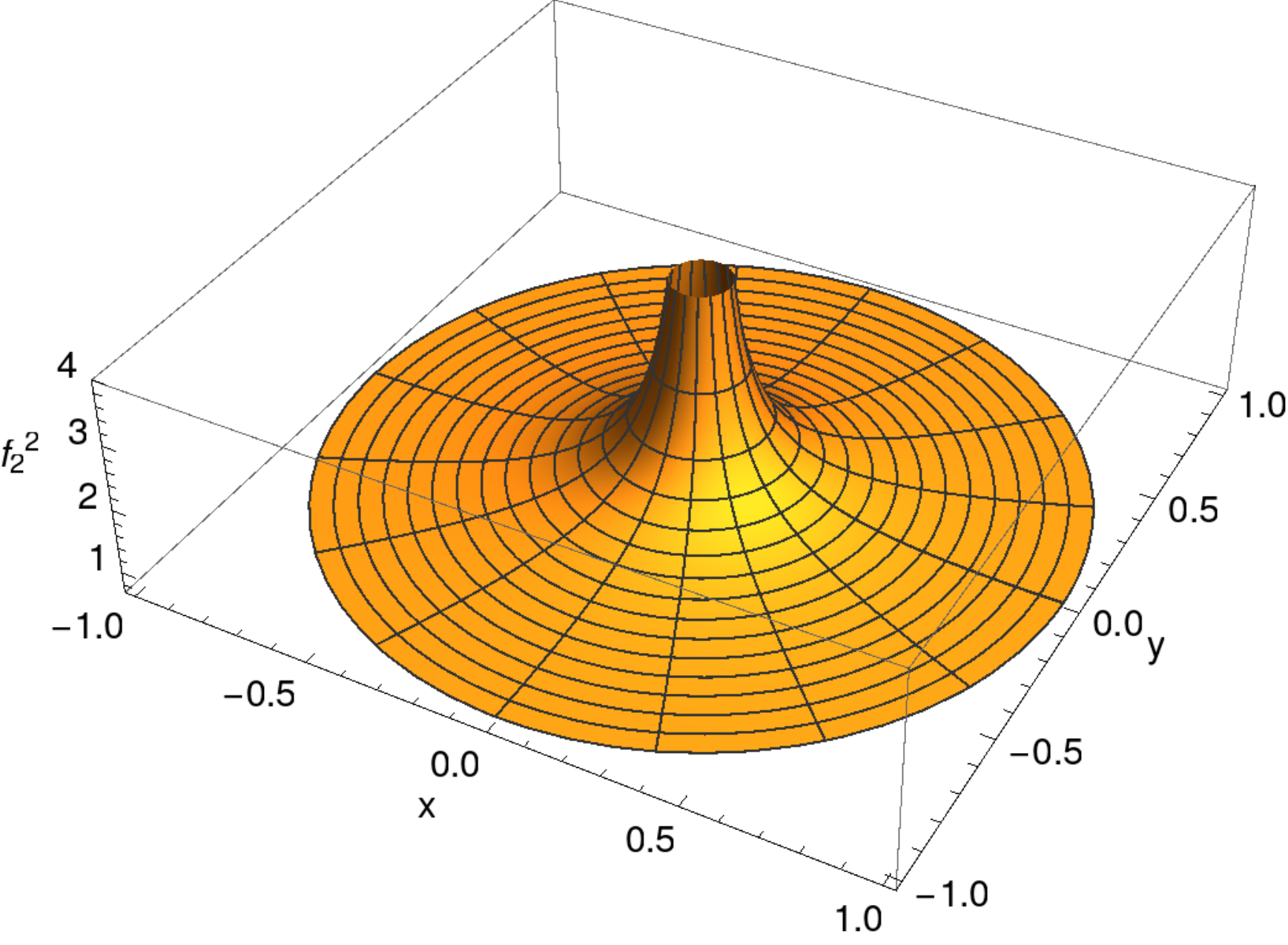}

\vskip 6mm
  \begin{tabular}{cc}
   \includegraphics[height=0.315\linewidth]{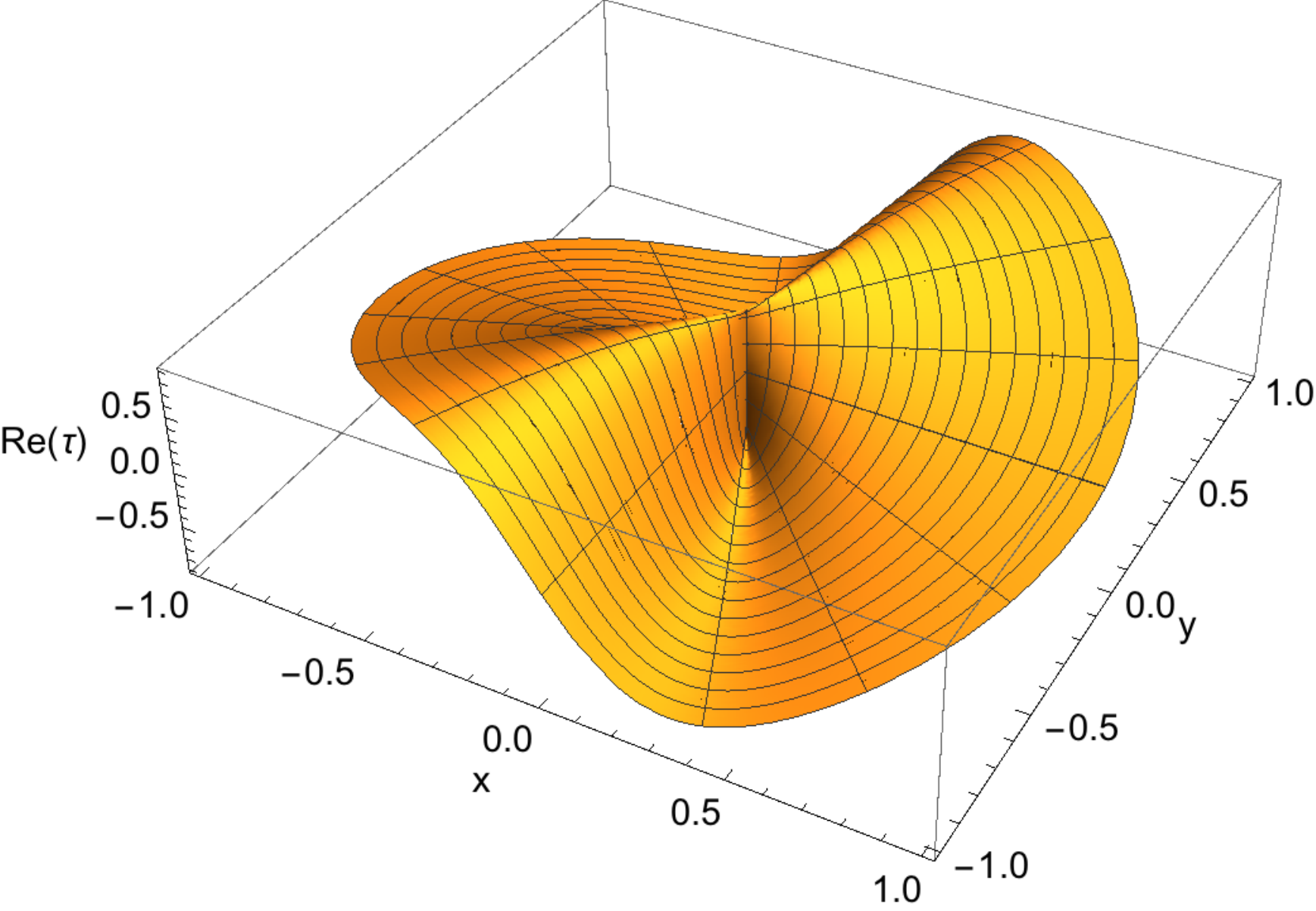}&\hskip 0.25in
   \includegraphics[height=0.315\linewidth]{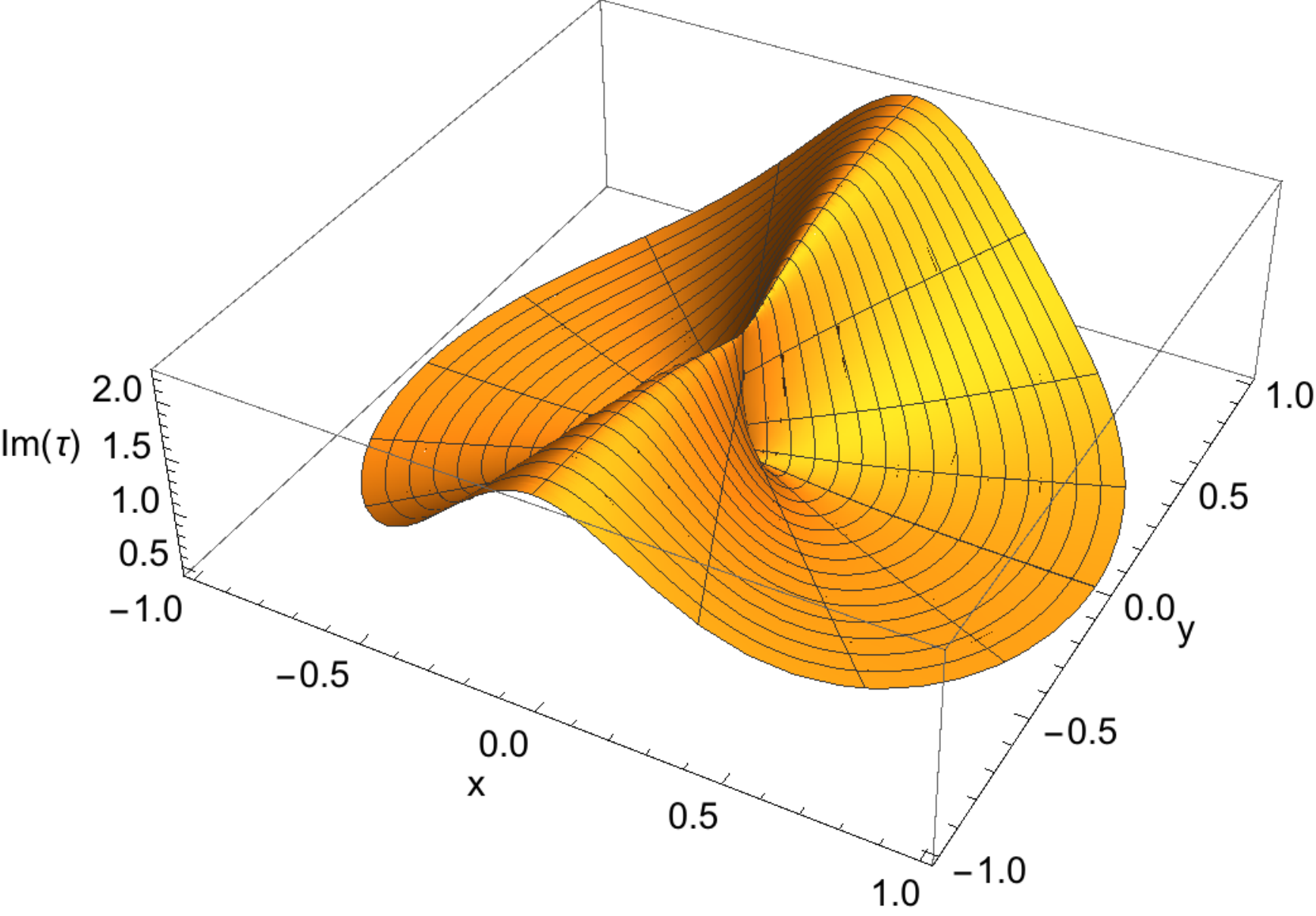}\\[8mm]
   \includegraphics[height=0.315\linewidth]{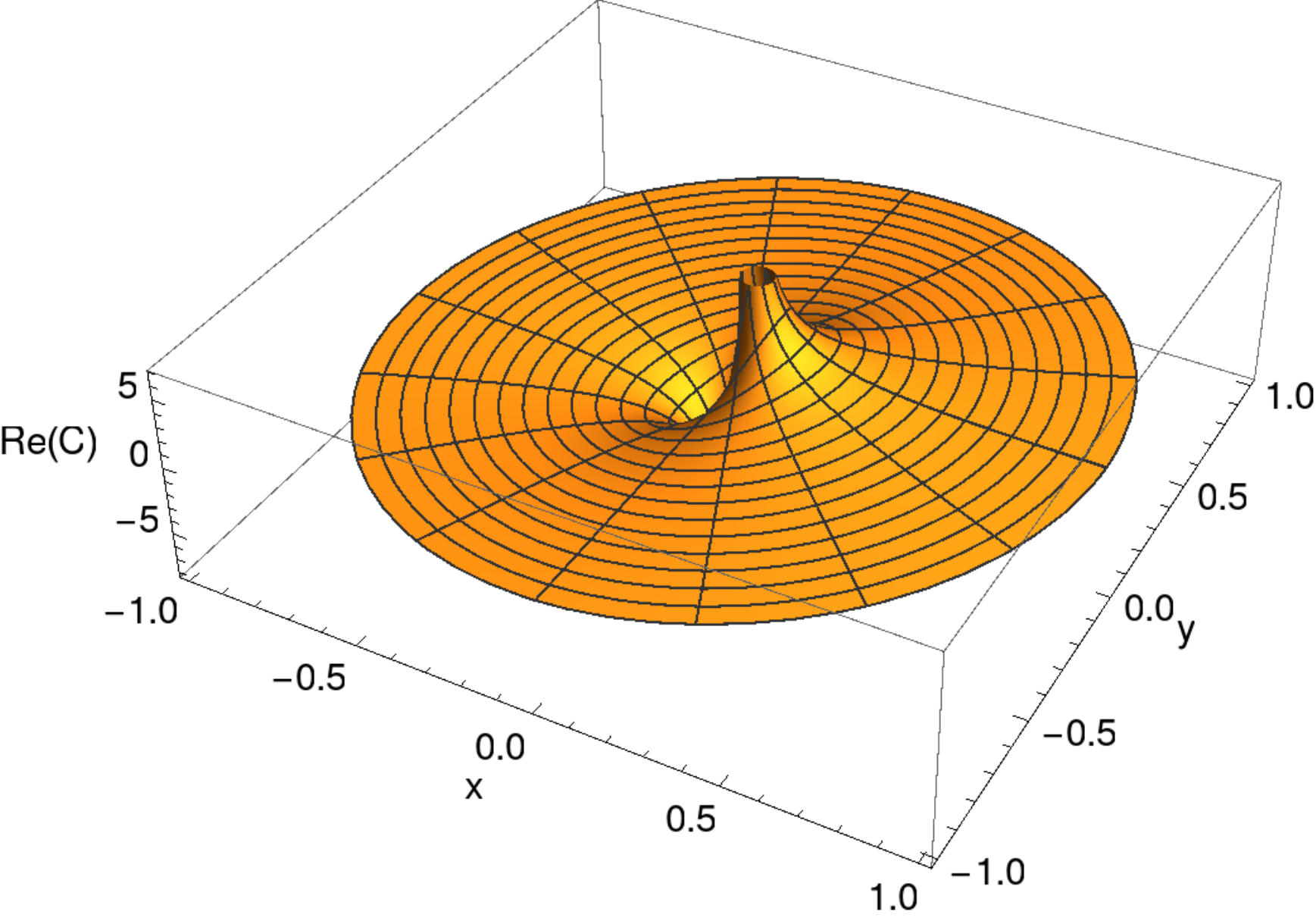}&\hskip 0.25in
   \includegraphics[height=0.315\linewidth]{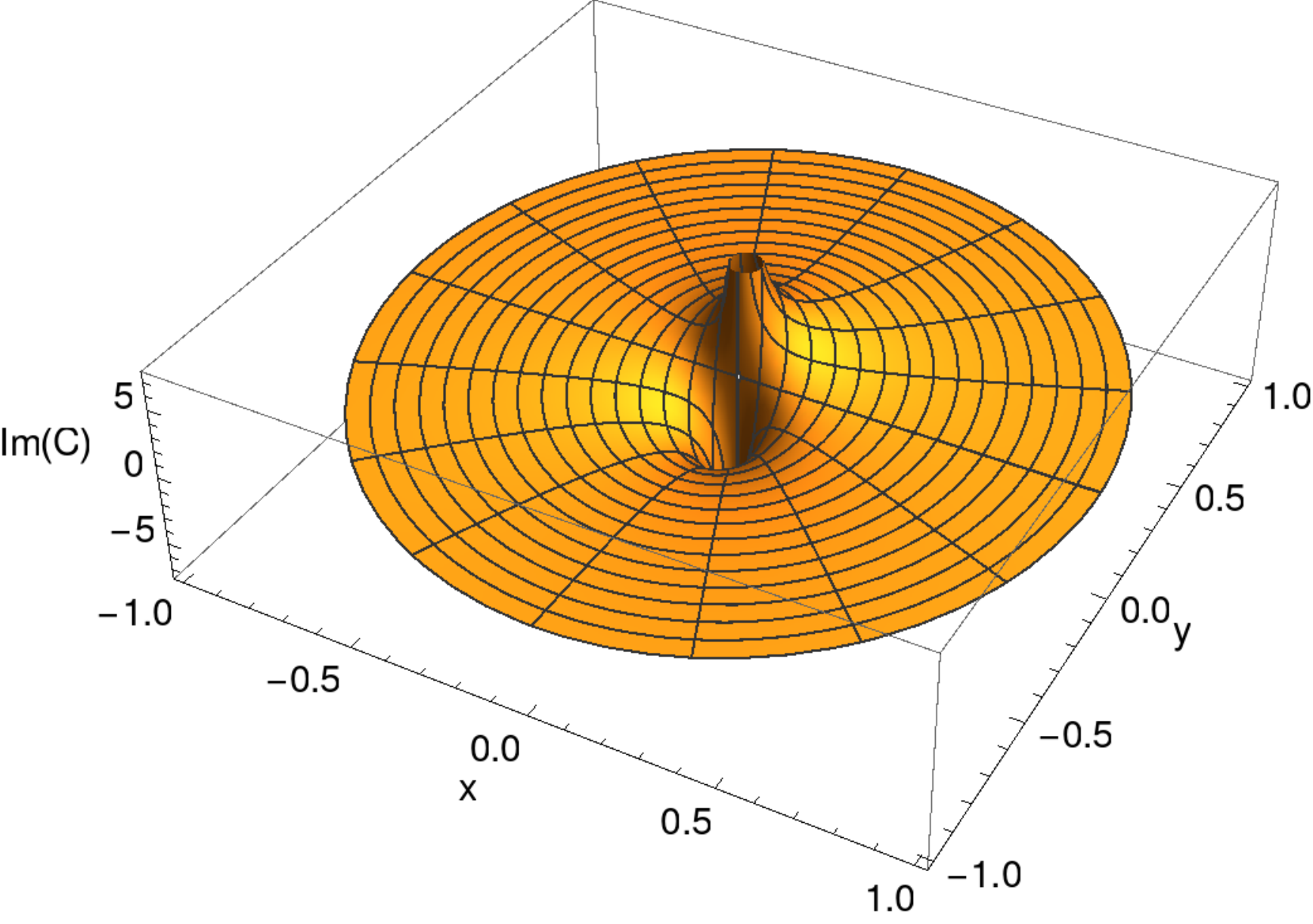}
  \end{tabular}
\caption{The metric factors, axion-dilaton, and two-form fields for the minimal solution with a single interior pole of order two. We have chosen $\varsigma = \tilde\cA_+^0={i \over 2}$ in (\ref{eq:cA-ex}). Because the metric factors all take the same qualitative form, we display only $f_2^2$. }
\label{fig:minimalplots}
\end{figure}

In Figure \ref{fig:minimalplots} we plot the metric factors, axion-dilaton, and two-form fields for a minimal solution on the disc. In order to obtain these plots, we have chosen $\varsigma = \tilde\cA_+^0={i \over 2}$ in (\ref{eq:cA-ex}). Because the metric factors all take the same qualitative form, we have only included the plot for $f_2^2$.

\subsection{Non-minimal solutions on the disc}

Generalizing to a single pole at $t=t_1$ of higher order $\nu = \nu_1>2$,  we have,  
\begin{align}
\label{generalinterior}
 \partial_w\cA_+&=\frac{\bar \sigma}{(w-\bar t)^{\nu}}\prod_{k=1}^{\nu-2} (w-u_k) 
 & \partial_w\cA_-&= \frac{\sigma}{(w-t)^{\nu}} \prod_{k=1}^{\nu-2} (w- \bar u_k)
\end{align}
with $\Im (u_k) >0$.  Since $\partial_w\cA_-$ has no zeros in the upper half-plane, $\kappa^2$ is positive throughout $\Sigma$. The pole at $t$ can be mapped to $w=i$ by $SL(2,\RR)$. By partial fraction decomposition the differentials can then be rewritten as,
\begin{align}
 \partial_w\cA_\pm&=\sum_{k=2}^{\nu}\frac{Z_{\pm}^{k}}{(w\pm i)^k}
 &
 Z_+^k&=\bar\sigma h_{\nu-2,k} & Z_-^k=\overline{Z_+^k}
\end{align}
where  $h_{\nu-2,k}(u_i)$ are $k$-th order symmetric polynomials in the $\nu-2$ zeroes $u_i$ given by 
\begin{align}
h_{\nu-2,k}(u_i) &= (-1)^k \sum_{\ell_1 < \dots < \ell_{k-2}}^{\nu-2}(u_{\ell_1} + i)\dots (u_{\ell_{k-2}}+i)
\end{align}
The residues of $\partial_w\cA_\pm$ at $w=\mp i$ vanish, such that $\cA_\pm$ are single-valued and given by,
\begin{align}
\label{eq:general-pole-disc}
 \cA_\pm&=\cA_\pm^0-\sum_{k=2}^\nu \frac{1}{k-1}\frac{Z_\pm^k}{(w\pm i)^{k-1}}
\end{align}
To ensure that $\cG$ is single-valued we compute $\partial_w\cB$,
\begin{align}
 \partial_w\cB&=\cA_+^0\partial_w\cA_- - \cA_-^0\partial_w\cA_+
 -\sum_{k,\ell=2}^\nu\frac{1}{k-1}\left[\frac{Z_+^kZ_-^\ell}{(w+i)^{k-1}(w-i)^\ell}-\frac{Z_-^kZ_+^\ell}{(w-i)^{k-1}(w+i)^\ell}\right]
\end{align}
The residue at $w=i$ is found to be 
\begin{align}
\mathrm{Res} (\p_w\cB) \Big |_{w=i} &=- \sum_{k,\ell=2}^\nu \frac{2^{2-k-\ell}}{k-1} \binom{k+\ell-3}{k-2} \left[i^{\ell-k} Z_+^kZ_-^\ell + (-i)^{\ell-k} Z_-^k Z_+^\ell \right]
\end{align}
This is real, satisfying (\ref{eq:Apm-res}) for any choice of the zeroes $u_i$. Thus, $\cG$ is single-valued for all differentials of the form (\ref{generalinterior}). It remains to implement $\cG<0$ and $0<T^2<1$. Explicit investigation shows that this is satisfied e.g.\ for $\cA_\pm^0=0$ and $u_i=\ldots=u_{\nu-2}=2i$ for $\nu \in\lbrace 3,4,5,6\rbrace$.

\begin{figure}
\centering
\includegraphics[height=0.315\linewidth]{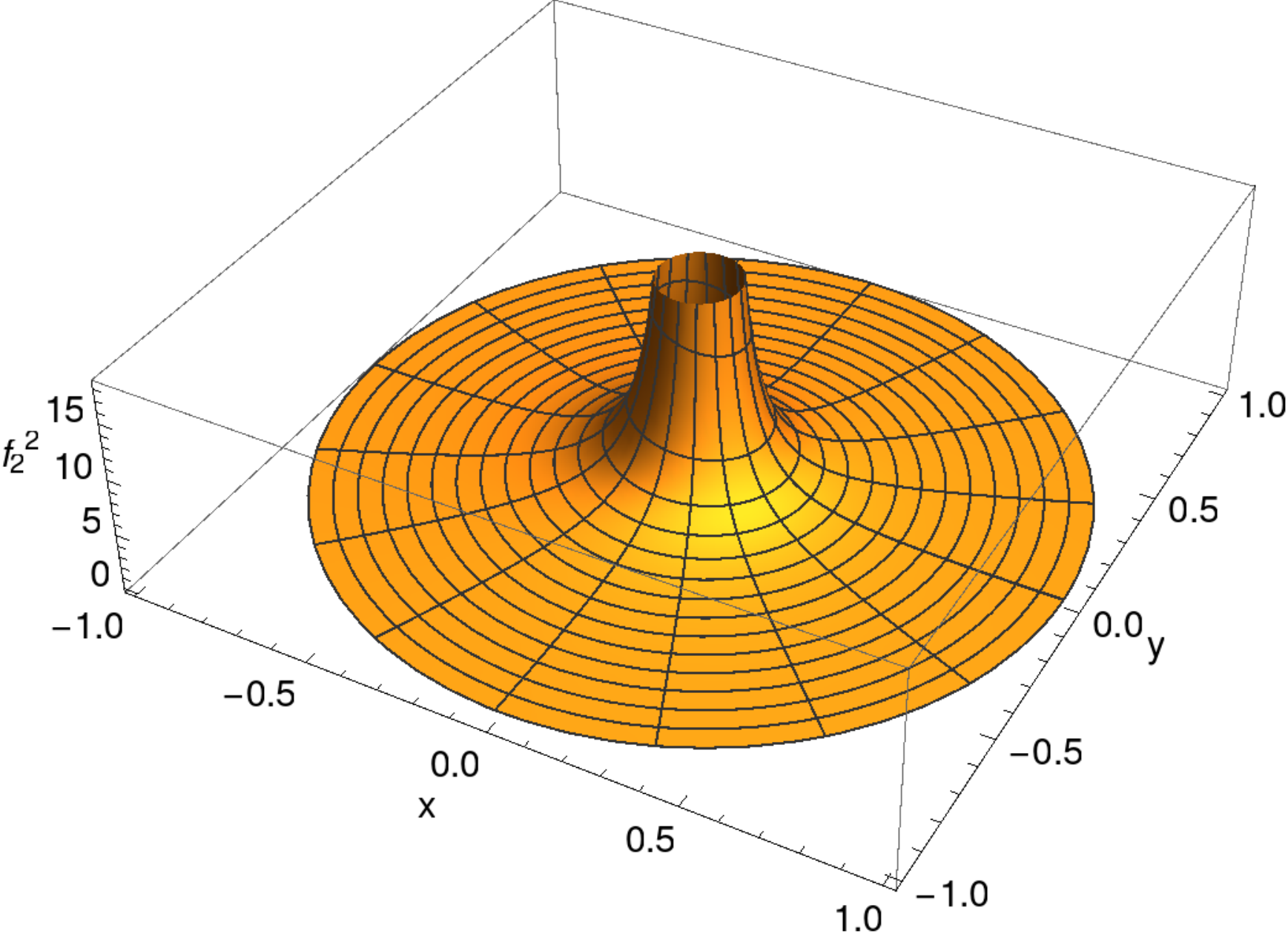}

\vskip 6mm
  \begin{tabular}{cc}
   \includegraphics[height=0.315\linewidth]{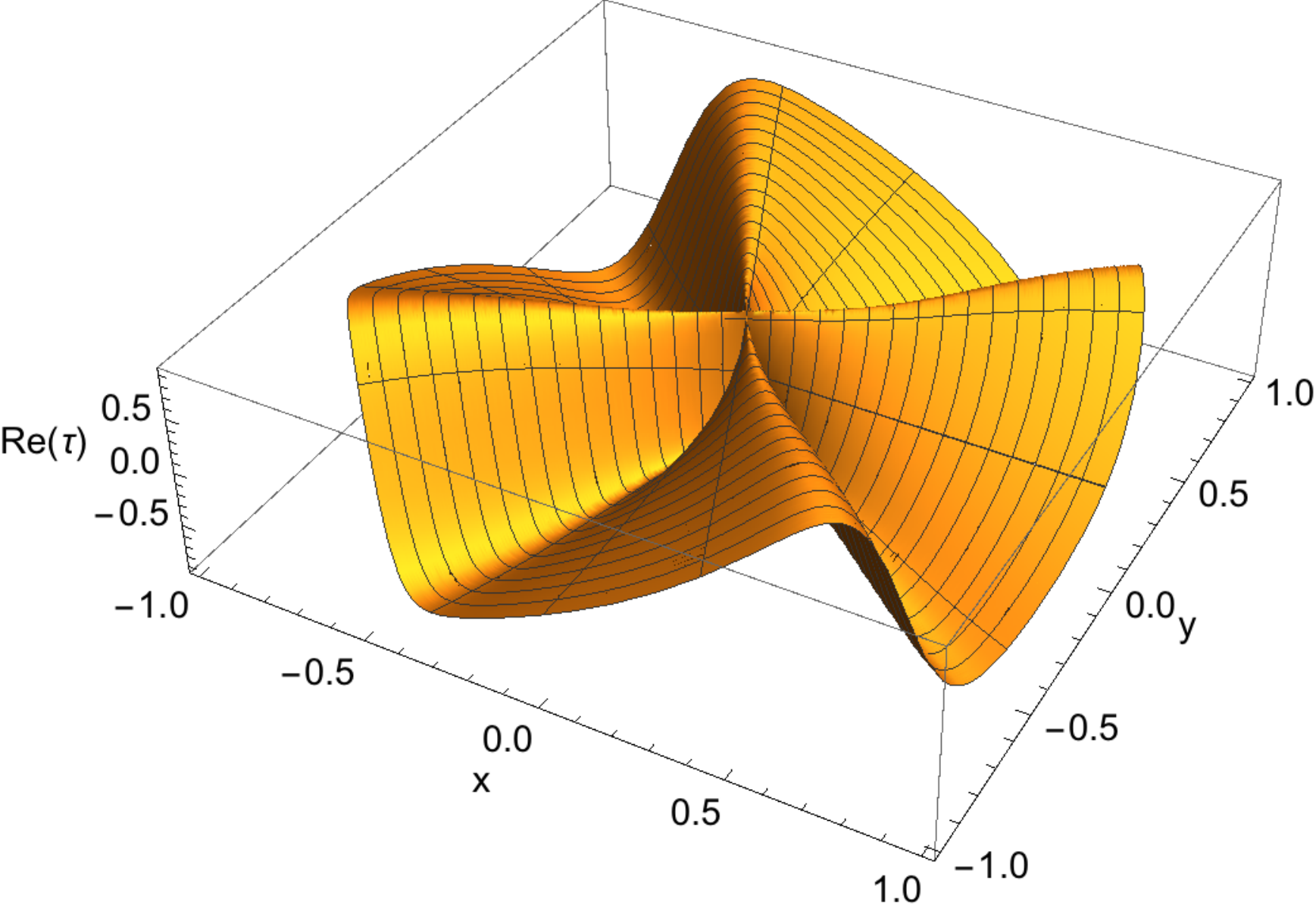}&\hskip 0.25in
   \includegraphics[height=0.315\linewidth]{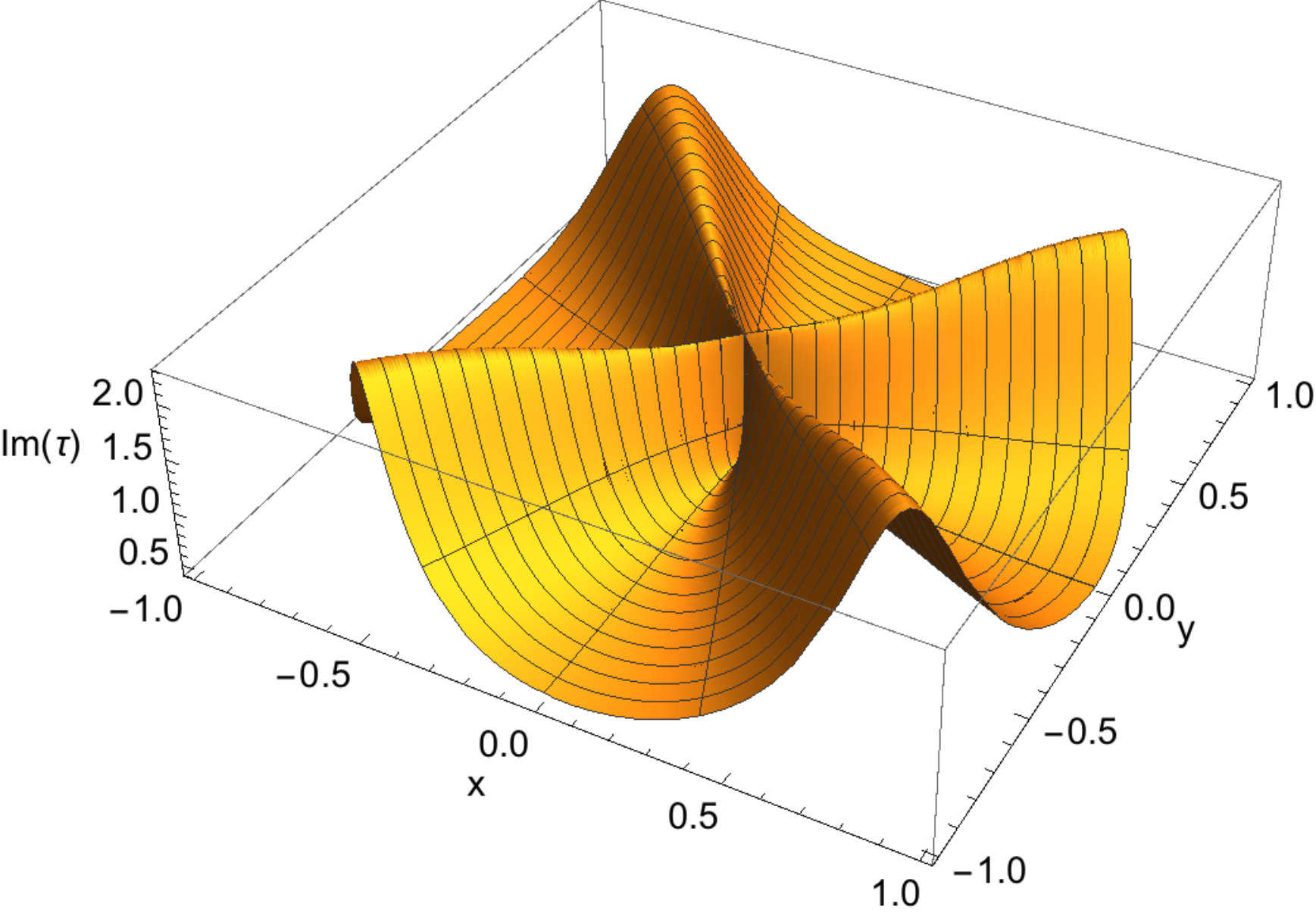}\\[8mm]
   \includegraphics[height=0.315\linewidth]{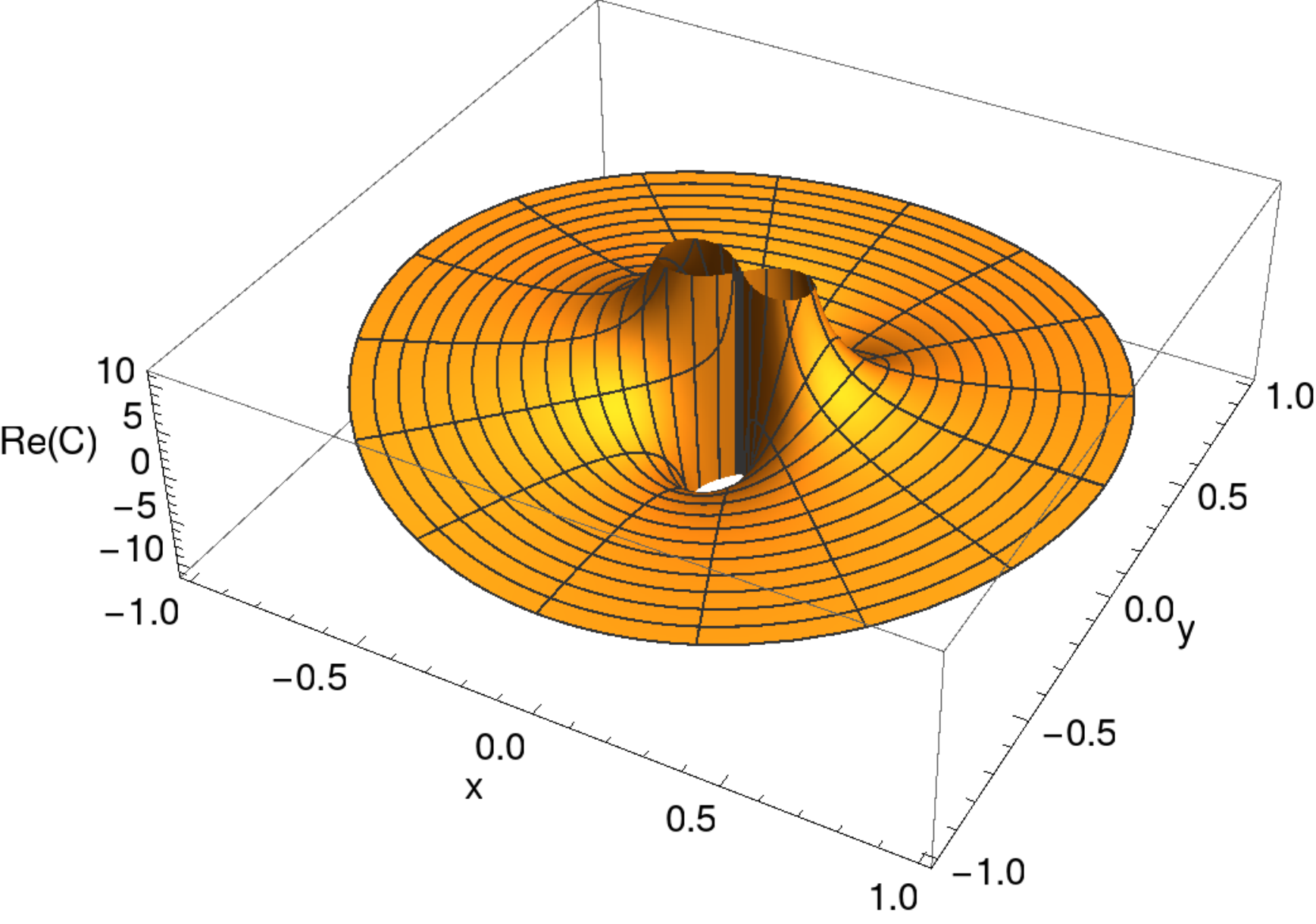}&\hskip 0.25in
   \includegraphics[height=0.315\linewidth]{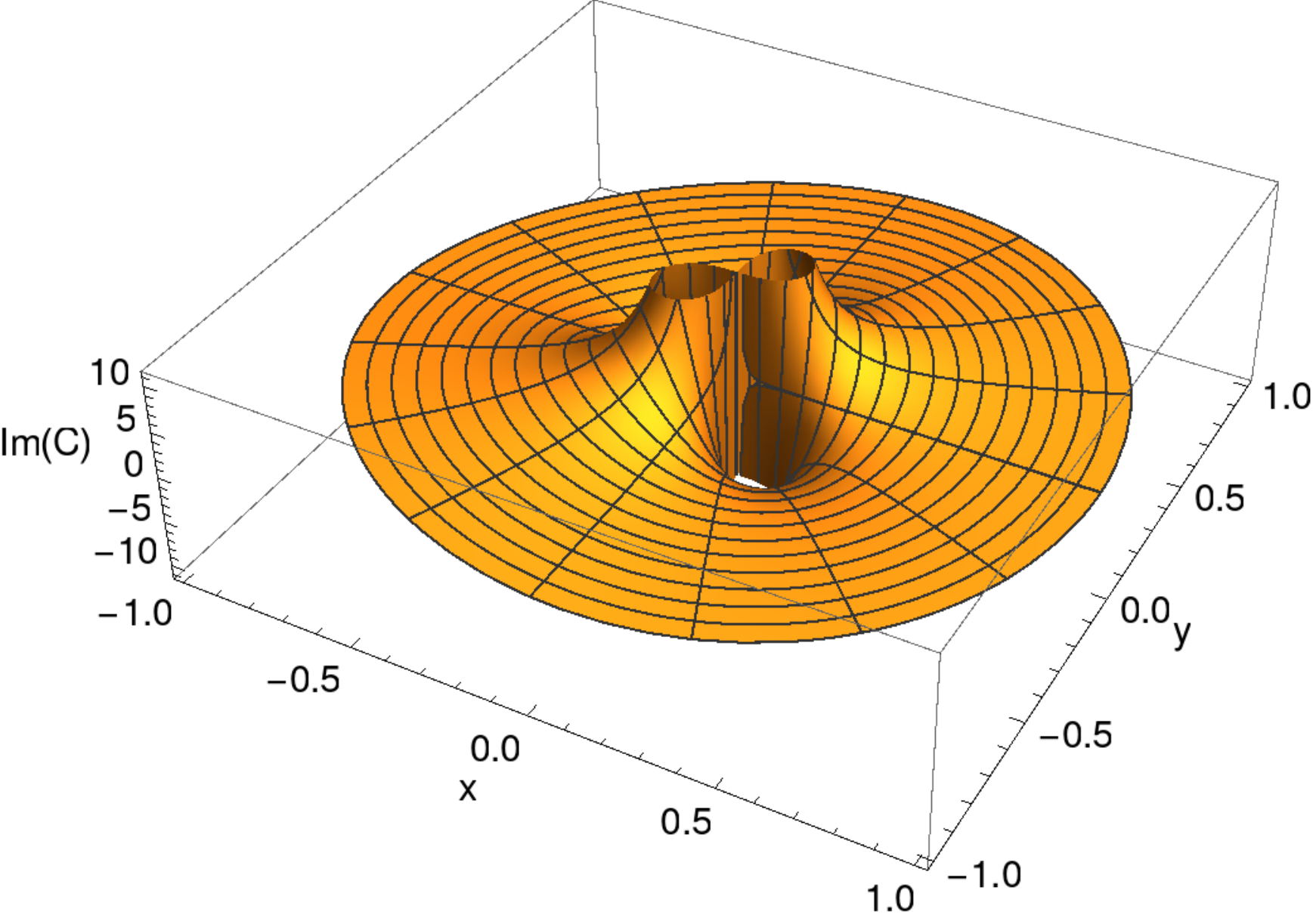}
  \end{tabular}
\caption{The metric factors, axion-dilaton, and two-form fields for the solution with a single interior pole of order three. We have chosen $\cA_\pm^0=0,\sigma=1,$ and $u_1 = 2 i$ in (\ref{eq:general-pole-disc}). Because the metric factors all take the same qualitative form, we display only $f_2^2$. }
\label{fig:n3plots}
\end{figure}

\sm

A more symmetric solution can be realized conveniently by working directly on the disc,  $\Sigma=\lbrace z\in\CC \,\big\vert\, |z|^2\leq 1\rbrace$, with the coordinate transformation used in (\ref{eq:cA-ex}) and the differentials
\begin{align}
 \partial_z\cA_+&=\bar\varsigma\left(z^{\nu-2}-\alpha\right) & \partial_z\cA_-&=-\frac{\varsigma}{z^2}\left(z^{2-\nu}-\alpha\right)
\end{align}
with $0<\alpha<1$. This corresponds to a pole at $z=0$ in $\partial_z\cA_-$ and zeros at the $(\nu-2)$\textsuperscript{th} roots of $\alpha$ in $\partial_z\cA_+$. The functions $\cA_\pm$ are
\begin{align}
\cA_+&=\cA_+^0+z\bar\varsigma\left(\frac{z^{\nu-2}}{\nu-1}-\alpha\right)
&
\cA_-&=\cA_-^0+\frac{\varsigma}{z}\left(\frac{z^{2-\nu}}{\nu-1}-\alpha\right)
\end{align}
They satisfy $\overline{\cA_\mp(1/\bar z)}=\cA_\pm$.  Under the $\ZZ_{\nu-2}$ generated by $z\rightarrow z\exp(\frac{2\pi i}{\nu-2})$ and an appropriate transformation of $\cA_\pm^0$, the functions $\cA_\pm$ transform by an overall multiplicative phase. These $\ZZ_{\nu-2}$ transformations leave $\kappa^2$, $\cG$, and $T$, and consequently the metric functions invariant. The RR potentials and axion-dilaton scalar transform non-trivially.

\sm

The general Ansatz for the differentials (\ref{generalinterior}) has $\nu$ complex parameters. Adding the integration constants $\cA^0_\pm$ and subtracting 3 real degrees of freedom for the $SL(2,\RR)$ automorphisms of the upper half-plane leaves $2\nu-1$ real parameters. These are subject to additional regularity constraints to implement $\cG<0$ and $0<T^2<1$. The examples given above show that these constraints can be satisfied. Since the additional regularity conditions take the form of inequalities, there is indeed a $2\nu-1$ parameter family of solutions for each $\nu$. The $SU(1,1)$ duality transformations (\ref{eq:SU11}) with $v\neq 0$ again do not map the set of functions (\ref{eq:general-pole-disc}) into itself, and instead carve out $SU(1,1)$ orbits of regular solutions. Adding two parameters for the $SU(1,1)$ transformations with $v\neq 0$ and subtracting one degree for the constant shifts which only produce gauge transformations of $\cC$ and $\cM$, we arrive at $2\nu$ parameters.

\sm

In Figure \ref{fig:n3plots}, we plot the metric factors, axion-dilaton, and two-form fields for the simplest non-minimal solution on the disc, i.e. the solution with $n=3$. To obtain these plots, we have chosen $\cA_\pm^0=0,\sigma=1,$ and $u_1 = 2 i$ in (\ref{eq:general-pole-disc}). As in the case of the minimal solutions, the metric factors all have the same qualitative form, and hence we include the result only for $f_2^2$.

%%%%%%%%%%%%%%%%%%%%%%%%%%%%%%%%%%%%%%%%%%%%%%%%%%%%%%
%%%%%%%%%%%%%%%%%%%%%%%%%%%%%%%%%%%%%%%%%%%%%%%%%%%%%%
\section{T-dual of \texorpdfstring{$AdS_2\times S^7$}{AdS2xS7} in Type IIA}
\label{app:T-dual}
%%%%%%%%%%%%%%%%%%%%%%%%%%%%%%%%%%%%%%%%%%%%%%%%%%%%%%
%%%%%%%%%%%%%%%%%%%%%%%%%%%%%%%%%%%%%%%%%%%%%%%%%%%%%%

In this section, we comment on possible T-duals of a class of $AdS_2$ solutions in Type IIA.  T-duals of a Type IIA solution with geometry $AdS_6$ warped over a half sphere $S^4$ describing the D4-D8 system could be recovered as special cases of the general local $AdS_6$ solution in Type IIB \cite{D'Hoker:2016rdq,Hong:2018amk,Lozano:2018pcp}.  T-dualizing the Type IIA solution along the $S^1$ Hopf fiber in $S^3$ produces a supersymmetric solution in Type IIB, as does non-Abelian T-duality \cite{Lozano:2012au}. With the $AdS_6$ superalgebra being unique, these T-duals had to be contained in the solutions of \cite{D'Hoker:2016rdq}.

\sm

For the case of $AdS_2$, the solutions of \cite{Dibitetto:2018gbk} describing semi-localized D0-D8-F1 systems in massive Type IIA take the form $AdS_2\times S^7$ warped over an interval.  One could consider various $U(1)$ isometries with fixed points in $S^7$ for T-duality. Recalling that $S^{2 n + 1}$ is a $U(1)$ bundle over $\CC\PP^n$, the $S^1$ fiber is a natural candidate on which to carry out this T-duality. Such T-dualities have been discussed in \cite{Duff:1998hj,Duff:1998us,Duff:1998cr}. With the $S^p\times S^q$ slicing of $S^n$, where $p+q=n-1$, the $S^1$ in the $S^1\times S^5$ slicing may be another candidate. The $S^1$ Hopf fiber in $S^3$ could be used in one of the factors of the $S^3\times S^3$ slicing, or in the Hopf fibration of $S^3$ over $S^4$. Non-Abelian T-duality may provide further options. None of these options, however, would produce an $S^6$ in the T-dual geometry. 
Moreover, even where a superalgebra with the preserved bosonic symmetries exists, supersymmetry may not be preserved by T-duality along $U(1)$ isometries with fixed points (see, however, \cite{Alvarez:1995np,Alvarez:1995ai}).

\sm

We discuss the case of $\CC\PP^3$ in the following. We illustrate the massless case, when there are no D8-branes, in the conventions of \cite{Dibitetto:2018gbk}, and consider the semi-localized intersection of D0 and F1. The generalization to the massive case is straightforward. The string-frame metric, dilaton, and two-form field are given by,
\begin{align}
ds_{10}^2 &= \frac{1}{4}L^2 W^2 \left[ ds_{AdS_2}^2 + 4 \left( d \theta^2 + 4 \sin^2 \theta ds_{S^7}^2 \right) \right]
\no\\
\Im(\tau)&= \frac{1}{g_s L W}
\hskip 20mm
B_{2} =- B_0 W^2 \cos\theta\, \mathrm{vol}_{AdS_2}
\end{align}
where $g_s>0$ and $B_0$ are constants, $\theta \in[0, \pi]$, $L^2 = \frac{1}{8}\sqrt{Q_{D0} Q_{F1}}$, and $W = (\sin \theta)^{-3/2}$.
The F1-string wraps a combination of the AdS$_2$ radial direction and the $\theta$-direction.
T-dualizing along the $S^1$ fiber in the fibration over $\CC\PP^3$ yields the string-frame metric and dilaton
\begin{align}
ds_{10}^2 &= {1 \over 4} L^2 W^2 \left[ ds_{AdS_2}^2 + 4 \sin^2 \theta \,ds^2_{\CC\PP^3} \right] + {4 \over L^2 W^2 \sin^2 \theta} \left[d \psi^2 + {1 \over 4} L^4 W^4 \sin^2 \theta d \theta^2 \right]
\no\\
\Im(\tau)&= \frac{1}{g_s}\sin\theta
\end{align}
Such a configuration could naturally arise from Type IIB solutions of the form $AdS_2\times \CC\PP^3$ warped over a Riemann surface $\Sigma$, where the appropriate superalgebra would be $SU(1,1\vert 4)$. These solutions are currently under investigation.

%%%%%%%%%%%%%%%%%%%%%%%%%%%%%%%%%%%%%%%%%%%%%%%%%%
%%%%%%%%%%%%%%%%%%%%%%%%%%%%%%%%%%%%%%%%%%%%%%%%%%
\section{Discussion}
\label{sec:discussion}
%%%%%%%%%%%%%%%%%%%%%%%%%%%%%%%%%%%%%%%%%%%%%%%%%%
%%%%%%%%%%%%%%%%%%%%%%%%%%%%%%%%%%%%%%%%%%%%%%%%%%

In the first part of this work we have constructed an Ansatz for global Type IIB supergravity solutions with 16 supersymmetries on a space-time of the from $AdS_2\times S^6$  warped over the unit disc or equivalently the upper half-plane, which may allow for an identification with string junctions.  These solutions circumvent  the no-go results of \cite{Corbino:2017tfl}, and naturally implement the boundary conditions on $\partial\Sigma$ which impose stronger constraints than in the $AdS_6\times S^2$ case. The remaining conditions for regularity and geodesic completeness were reduced to algebraic constraints on the parameters of the Ansatz, whose complete solution remains an open problem.  In analogy with the relation of $AdS_6$ solutions to M5-brane curves \cite{Kaidi:2018zkx}, one may expect the data $(\Sigma,\cA_\pm)$ for solutions corresponding to string junctions to define the  holomorphic curve wrapped by the M2-brane in the M-theory uplift of the string junctions \cite{Matsuo:1997jw,Krogh:1997dx,Rey:1997sp,Shocklee:2001df}. As discussed in sec.~\ref{app:T-dual}, the T-duals of Type IIA $AdS_2$ solutions relating to D0-F1-D8 systems naturally realize a different superalgebra, motivating their further investigation.

\sm

In the second part we studied solutions independently from a string junction interpretation. We presented families of non-compact solutions with geometry $AdS_2\times S^6$ warped over a punctured sphere or a punctured disc. At the punctures the geometry decompactifies into an asymptotic region. The explicit solutions we have presented are infinite families which all have one asymptotic region.  They are labeled by an integer $\nu$, which corresponds to a $\ZZ_\nu$ symmetry in the asymptotic region, and have $2\nu$ real parameters.

\sm

We laid out a systematic construction strategy which may give access to further solutions with similar features.
Possible generalizations include solutions with multiple asymptotic regions or Riemann surfaces of different topology.
The construction of these $AdS_2$ solutions involves poles of the functions $\cA_\pm$ in the interior of $\Sigma$, which would not be compatible with the regularity conditions in the $AdS_6$ case \cite{DHoker:2017mds}. This highlights the physical independence of these two cases, despite the similarities in the general local solution to the BPS equations.  The $AdS_2$ solutions share certain features with black hole micro-state geometries \cite{Lunin:2001jy,Giusto:2004id,Bena:2015bea,Bena:2016ypk,Bena:2017xbt,Bena:2018bbd}, but we leave their interpretation for future work. 

\sm

Various further generalizations of the solutions may be possible.
For example, one may try to combine the two elements discussed in this paper - namely, to realize the singularities allowing for a local match to $(p,q)$-string solutions discussed in the first part in the non-compact solutions with asymptotic regions discussed in the second part.
The $AdS_6$ solutions can be generalized to include 7-branes by modifying the functions $\cA_\pm$ of a regular solution \cite{DHoker:2017zwj}, and it may be possible to generalize the $AdS_2$ solutions in a similar way. Further solutions may also be constructed by relaxing the regularity conditions, to allow e.g.\ for smeared branes along the lines of \cite{Lozano:2018pcp}.

\section*{Acknowledgements}

We thank Juan Maldacena, Ben Michel, and Moshe Rozali for very interesting discussions.
This research is supported in part by the National Science Foundation (NSF) under grant PHY-16-19926.
David Corbino, Justin Kaidi, and Christoph Uhlemann gratefully acknowledge support from the Mani L. Bhaumik Institute for Theoretical Physics at UCLA.

\bibliographystyle{JHEP.bst}
\bibliography{ads2glob}

\providecommand{\href}[2]{#2}\begingroup\raggedright\begin{thebibliography}{10}

\bibitem{Maldacena:1997re}
J.~M. Maldacena, \emph{{The Large N limit of superconformal field theories and
  supergravity}}, \href{https://doi.org/10.1023/A:1026654312961,
  10.4310/ATMP.1998.v2.n2.a1}{\emph{Int. J. Theor. Phys.} {\bfseries 38} (1999)
  1113--1133}, [\href{https://arxiv.org/abs/hep-th/9711200}{{\ttfamily
  hep-th/9711200}}].

\bibitem{DHoker:2008wvd}
E.~D'Hoker, J.~Estes, M.~Gutperle, D.~Krym and P.~Sorba, \emph{{Half-BPS
  supergravity solutions and superalgebras}},
  \href{https://doi.org/10.1088/1126-6708/2008/12/047}{\emph{JHEP} {\bfseries
  12} (2008) 047}, [\href{https://arxiv.org/abs/0810.1484}{{\ttfamily
  0810.1484}}].

\bibitem{Brandhuber:1999np}
A.~Brandhuber and Y.~Oz, \emph{{The D-4 - D-8 brane system and five-dimensional
  fixed points}},
  \href{https://doi.org/10.1016/S0370-2693(99)00763-7}{\emph{Phys. Lett.}
  {\bfseries B460} (1999) 307--312},
  [\href{https://arxiv.org/abs/hep-th/9905148}{{\ttfamily hep-th/9905148}}].

\bibitem{Apruzzi:2014qva}
F.~Apruzzi, M.~Fazzi, A.~Passias, D.~Rosa and A.~Tomasiello, \emph{{AdS$_{6}$
  solutions of type II supergravity}},
  \href{https://doi.org/10.1007/JHEP11(2014)099,
  10.1007/JHEP05(2015)012}{\emph{JHEP} {\bfseries 11} (2014) 099},
  [\href{https://arxiv.org/abs/1406.0852}{{\ttfamily 1406.0852}}].

\bibitem{Kim:2015hya}
H.~Kim, N.~Kim and M.~Suh, \emph{{Supersymmetric AdS$_6$ Solutions of Type IIB
  Supergravity}},
  \href{https://doi.org/10.1140/epjc/s10052-015-3705-1}{\emph{Eur. Phys. J.}
  {\bfseries C75} (2015) 484},
  [\href{https://arxiv.org/abs/1506.05480}{{\ttfamily 1506.05480}}].

\bibitem{D'Hoker:2016rdq}
E.~D'Hoker, M.~Gutperle, A.~Karch and C.~F. Uhlemann, \emph{{Warped
  $AdS_6\times S^2$ in Type IIB supergravity I: Local solutions}},
  \href{https://doi.org/10.1007/JHEP08(2016)046}{\emph{JHEP} {\bfseries 08}
  (2016) 046}, [\href{https://arxiv.org/abs/1606.01254}{{\ttfamily
  1606.01254}}].

\bibitem{DHoker:2016ysh}
E.~D'Hoker, M.~Gutperle and C.~F. Uhlemann, \emph{{Holographic duals for
  five-dimensional superconformal quantum field theories}},
  \href{https://doi.org/10.1103/PhysRevLett.118.101601}{\emph{Phys. Rev. Lett.}
  {\bfseries 118} (2017) 101601},
  [\href{https://arxiv.org/abs/1611.09411}{{\ttfamily 1611.09411}}].

\bibitem{DHoker:2017mds}
E.~D'Hoker, M.~Gutperle and C.~F. Uhlemann, \emph{{Warped $AdS_6\times S^2$ in
  Type IIB supergravity II: Global solutions and five-brane webs}},
  \href{https://doi.org/10.1007/JHEP05(2017)131}{\emph{JHEP} {\bfseries 05}
  (2017) 131}, [\href{https://arxiv.org/abs/1703.08186}{{\ttfamily
  1703.08186}}].

\bibitem{DHoker:2017zwj}
E.~D'Hoker, M.~Gutperle and C.~F. Uhlemann, \emph{{Warped $AdS_6\times S^2$ in
  Type IIB supergravity III: Global solutions with seven-branes}},
  \href{https://doi.org/10.1007/JHEP11(2017)200}{\emph{JHEP} {\bfseries 11}
  (2017) 200}, [\href{https://arxiv.org/abs/1706.00433}{{\ttfamily
  1706.00433}}].

\bibitem{Corbino:2017tfl}
D.~Corbino, E.~D'Hoker and C.~F. Uhlemann, \emph{{AdS$_{2}$ $\times$ S$^{6}$
  versus AdS$_{6}$ $\times$ S$^{2}$ in Type IIB supergravity}},
  \href{https://doi.org/10.1007/JHEP03(2018)120}{\emph{JHEP} {\bfseries 03}
  (2018) 120}, [\href{https://arxiv.org/abs/1712.04463}{{\ttfamily
  1712.04463}}].

\bibitem{Sen:1997xi}
A.~Sen, \emph{{String network}},
  \href{https://doi.org/10.1088/1126-6708/1998/03/005}{\emph{JHEP} {\bfseries
  03} (1998) 005}, [\href{https://arxiv.org/abs/hep-th/9711130}{{\ttfamily
  hep-th/9711130}}].

\bibitem{Bergman:1998gs}
O.~Bergman and B.~Kol, \emph{{String webs and 1/4 BPS monopoles}},
  \href{https://doi.org/10.1016/S0550-3213(98)00565-3}{\emph{Nucl. Phys.}
  {\bfseries B536} (1998) 149--174},
  [\href{https://arxiv.org/abs/hep-th/9804160}{{\ttfamily hep-th/9804160}}].

\bibitem{Lunin:2008tf}
O.~Lunin, \emph{{Brane webs and 1/4-BPS geometries}},
  \href{https://doi.org/10.1088/1126-6708/2008/09/028}{\emph{JHEP} {\bfseries
  09} (2008) 028}, [\href{https://arxiv.org/abs/0802.0735}{{\ttfamily
  0802.0735}}].

\bibitem{VanProeyen:1986me}
A.~Van~Proeyen, \emph{{SUPERCONFORMAL ALGEBRAS}},  in \emph{{IN *VANCOUVER
  1986, PROCEEDINGS, SUPER FIELD THEORIES* 547-555.}}, 1986.

\bibitem{DHoker:2007mci}
E.~D'Hoker, J.~Estes and M.~Gutperle, \emph{{Gravity duals of half-BPS Wilson
  loops}}, \href{https://doi.org/10.1088/1126-6708/2007/06/063}{\emph{JHEP}
  {\bfseries 06} (2007) 063},
  [\href{https://arxiv.org/abs/0705.1004}{{\ttfamily 0705.1004}}].

\bibitem{Kim:2006qu}
N.~Kim and J.-D. Park, \emph{{Comments on AdS(2) solutions of D=11
  supergravity}},
  \href{https://doi.org/10.1088/1126-6708/2006/09/041}{\emph{JHEP} {\bfseries
  09} (2006) 041}, [\href{https://arxiv.org/abs/hep-th/0607093}{{\ttfamily
  hep-th/0607093}}].

\bibitem{Hartman:2008dq}
T.~Hartman and A.~Strominger, \emph{{Central Charge for AdS(2) Quantum
  Gravity}}, \href{https://doi.org/10.1088/1126-6708/2009/04/026}{\emph{JHEP}
  {\bfseries 04} (2009) 026},
  [\href{https://arxiv.org/abs/0803.3621}{{\ttfamily 0803.3621}}].

\bibitem{Almheiri:2014cka}
A.~Almheiri and J.~Polchinski, \emph{{Models of AdS$_{2}$ backreaction and
  holography}}, \href{https://doi.org/10.1007/JHEP11(2015)014}{\emph{JHEP}
  {\bfseries 11} (2015) 014},
  [\href{https://arxiv.org/abs/1402.6334}{{\ttfamily 1402.6334}}].

\bibitem{Jensen:2016pah}
K.~Jensen, \emph{{Chaos in AdS$_2$ Holography}},
  \href{https://doi.org/10.1103/PhysRevLett.117.111601}{\emph{Phys. Rev. Lett.}
  {\bfseries 117} (2016) 111601},
  [\href{https://arxiv.org/abs/1605.06098}{{\ttfamily 1605.06098}}].

\bibitem{Engelsoy:2016xyb}
J.~Engels, T.~G. Mertens and H.~Verlinde, \emph{{An investigation of AdS$_{2}$
  backreaction and holography}},
  \href{https://doi.org/10.1007/JHEP07(2016)139}{\emph{JHEP} {\bfseries 07}
  (2016) 139}, [\href{https://arxiv.org/abs/1606.03438}{{\ttfamily
  1606.03438}}].

\bibitem{Maldacena:2016upp}
J.~Maldacena, D.~Stanford and Z.~Yang, \emph{{Conformal symmetry and its
  breaking in two dimensional Nearly Anti-de-Sitter space}},
  \href{https://doi.org/10.1093/ptep/ptw124}{\emph{PTEP} {\bfseries 2016}
  (2016) 12C104}, [\href{https://arxiv.org/abs/1606.01857}{{\ttfamily
  1606.01857}}].

\bibitem{Suh:2018tul}
M.~Suh, \emph{{D4-branes wrapped on supersymmetric four-cycles}},
  \href{https://arxiv.org/abs/1809.03517}{{\ttfamily 1809.03517}}.

\bibitem{Hosseini:2018uzp}
S.~M. Hosseini, I.~Yaakov and A.~Zaffaroni, \emph{{Topologically twisted
  indices in five dimensions and holography}},
  \href{https://doi.org/10.1007/JHEP11(2018)119}{\emph{JHEP} {\bfseries 11}
  (2018) 119}, [\href{https://arxiv.org/abs/1808.06626}{{\ttfamily
  1808.06626}}].

\bibitem{Hosseini:2018usu}
S.~M. Hosseini, K.~Hristov, A.~Passias and A.~Zaffaroni, \emph{{6D attractors
  and black hole microstates}},
  \href{https://arxiv.org/abs/1809.10685}{{\ttfamily 1809.10685}}.

\bibitem{Suh:2018szn}
M.~Suh, \emph{{D4-branes wrapped on supersymmetric four-cycles from matter
  coupled $F(4)$ gauged supergravity}},
  \href{https://arxiv.org/abs/1810.00675}{{\ttfamily 1810.00675}}.

\bibitem{Massar:1999sb}
M.~Massar and J.~Troost, \emph{{D0 - D8 - F1 in massive IIA SUGRA}},
  \href{https://doi.org/10.1016/S0370-2693(99)00660-7}{\emph{Phys. Lett.}
  {\bfseries B458} (1999) 283--287},
  [\href{https://arxiv.org/abs/hep-th/9901136}{{\ttfamily hep-th/9901136}}].

\bibitem{Cvetic:2000cj}
M.~Cvetic, H.~Lu, C.~N. Pope and J.~F. Vazquez-Poritz, \emph{{AdS in warped
  space-times}}, \href{https://doi.org/10.1103/PhysRevD.62.122003}{\emph{Phys.
  Rev.} {\bfseries D62} (2000) 122003},
  [\href{https://arxiv.org/abs/hep-th/0005246}{{\ttfamily hep-th/0005246}}].

\bibitem{Imamura:2001cr}
Y.~Imamura, \emph{{1/4 BPS solutions in massive IIA supergravity}},
  \href{https://doi.org/10.1143/PTP.106.653}{\emph{Prog. Theor. Phys.}
  {\bfseries 106} (2001) 653--670},
  [\href{https://arxiv.org/abs/hep-th/0105263}{{\ttfamily hep-th/0105263}}].

\bibitem{Donos:2008ug}
A.~Donos, J.~P. Gauntlett and N.~Kim, \emph{{AdS Solutions Through
  Transgression}},
  \href{https://doi.org/10.1088/1126-6708/2008/09/021}{\emph{JHEP} {\bfseries
  09} (2008) 021}, [\href{https://arxiv.org/abs/0807.4375}{{\ttfamily
  0807.4375}}].

\bibitem{Dibitetto:2018gbk}
G.~Dibitetto and A.~Passias, \emph{{AdS$_{2}$ S$^{7}$ solutions from D0-F1-D8
  intersections}}, \href{https://doi.org/10.1007/JHEP10(2018)190}{\emph{JHEP}
  {\bfseries 10} (2018) 190},
  [\href{https://arxiv.org/abs/1807.00555}{{\ttfamily 1807.00555}}].

\bibitem{Dibitetto:2018gtk}
G.~Dibitetto and N.~Petri, \emph{{AdS$_{\textbf{2}}$ solutions and their
  massive IIA origin}},  \href{https://arxiv.org/abs/1811.11572}{{\ttfamily
  1811.11572}}.

\bibitem{Hong:2018amk}
J.~Hong, J.~T. Liu and D.~R. Mayerson, \emph{{Gauged Six-Dimensional
  Supergravity from Warped IIB Reductions}},
  \href{https://doi.org/10.1007/JHEP09(2018)140}{\emph{JHEP} {\bfseries 09}
  (2018) 140}, [\href{https://arxiv.org/abs/1808.04301}{{\ttfamily
  1808.04301}}].

\bibitem{Lozano:2018pcp}
Y.~Lozano, N.~T. Macpherson and J.~Montero, \emph{{$AdS_6$ T-duals and Type IIB
  $AdS_6\times S^2$ Geometries with 7-Branes}},
  \href{https://arxiv.org/abs/1810.08093}{{\ttfamily 1810.08093}}.

\bibitem{Schwarz:1995dk}
J.~H. Schwarz, \emph{{An SL(2,Z) multiplet of type IIB superstrings}},
  \href{https://doi.org/10.1016/0370-2693(95)01405-5,
  10.1016/0370-2693(95)01138-G}{\emph{Phys. Lett.} {\bfseries B360} (1995)
  13--18}, [\href{https://arxiv.org/abs/hep-th/9508143}{{\ttfamily
  hep-th/9508143}}].

\bibitem{Lozano:2012au}
Y.~Lozano, E.~\'O~Colg\'ain, D.~Rodr\'iguez-G\'omez and K.~Sfetsos,
  \emph{{Supersymmetric $AdS_6$ via T Duality}},
  \href{https://doi.org/10.1103/PhysRevLett.110.231601}{\emph{Phys. Rev. Lett.}
  {\bfseries 110} (2013) 231601},
  [\href{https://arxiv.org/abs/1212.1043}{{\ttfamily 1212.1043}}].

\bibitem{Duff:1998hj}
M.~J. Duff, \emph{{Anti-de Sitter space, branes, singletons, superconformal
  field theories and all that}},
  \href{https://doi.org/10.1142/S0217751X99000403}{\emph{Int. J. Mod. Phys.}
  {\bfseries A14} (1999) 815--844},
  [\href{https://arxiv.org/abs/hep-th/9808100}{{\ttfamily hep-th/9808100}}].

\bibitem{Duff:1998us}
M.~J. Duff, H.~Lu and C.~N. Pope, \emph{{AdS(5) x S**5 untwisted}},
  \href{https://doi.org/10.1016/S0550-3213(98)00464-7}{\emph{Nucl. Phys.}
  {\bfseries B532} (1998) 181--209},
  [\href{https://arxiv.org/abs/hep-th/9803061}{{\ttfamily hep-th/9803061}}].

\bibitem{Duff:1998cr}
M.~J. Duff, H.~Lu and C.~N. Pope, \emph{{AdS(3) x S**3 (un)twisted and
  squashed, and an O(2,2,Z) multiplet of dyonic strings}},
  \href{https://doi.org/10.1016/S0550-3213(98)00810-4}{\emph{Nucl. Phys.}
  {\bfseries B544} (1999) 145--180},
  [\href{https://arxiv.org/abs/hep-th/9807173}{{\ttfamily hep-th/9807173}}].

\bibitem{Alvarez:1995np}
E.~Alvarez, L.~Alvarez-Gaume and I.~Bakas, \emph{{T duality and space-time
  supersymmetry}},
  \href{https://doi.org/10.1016/0550-3213(95)00566-8}{\emph{Nucl. Phys.}
  {\bfseries B457} (1995) 3--26},
  [\href{https://arxiv.org/abs/hep-th/9507112}{{\ttfamily hep-th/9507112}}].

\bibitem{Alvarez:1995ai}
E.~Alvarez, L.~Alvarez-Gaume and I.~Bakas, \emph{{Supersymmetry and
  dualities}}, \href{https://doi.org/10.1016/0920-5632(96)00003-5}{\emph{Nucl.
  Phys. Proc. Suppl.} {\bfseries 46} (1996) 16--29},
  [\href{https://arxiv.org/abs/hep-th/9510028}{{\ttfamily hep-th/9510028}}].

\bibitem{Kaidi:2018zkx}
J.~Kaidi and C.~F. Uhlemann, \emph{{M-theory curves from warped AdS$_{6}$ in
  Type IIB}}, \href{https://doi.org/10.1007/JHEP11(2018)175}{\emph{JHEP}
  {\bfseries 11} (2018) 175},
  [\href{https://arxiv.org/abs/1809.10162}{{\ttfamily 1809.10162}}].

\bibitem{Matsuo:1997jw}
Y.~Matsuo and K.~Okuyama, \emph{{BPS condition of string junction from M
  theory}}, \href{https://doi.org/10.1016/S0370-2693(98)00288-3}{\emph{Phys.
  Lett.} {\bfseries B426} (1998) 294--296},
  [\href{https://arxiv.org/abs/hep-th/9712070}{{\ttfamily hep-th/9712070}}].

\bibitem{Krogh:1997dx}
M.~Krogh and S.~Lee, \emph{{String network from M theory}},
  \href{https://doi.org/10.1016/S0550-3213(98)00062-5}{\emph{Nucl. Phys.}
  {\bfseries B516} (1998) 241--254},
  [\href{https://arxiv.org/abs/hep-th/9712050}{{\ttfamily hep-th/9712050}}].

\bibitem{Rey:1997sp}
S.-J. Rey and J.-T. Yee, \emph{{BPS dynamics of triple (p, q) string
  junction}}, \href{https://doi.org/10.1016/S0550-3213(98)00401-5}{\emph{Nucl.
  Phys.} {\bfseries B526} (1998) 229--240},
  [\href{https://arxiv.org/abs/hep-th/9711202}{{\ttfamily hep-th/9711202}}].

\bibitem{Shocklee:2001df}
P.~Shocklee and L.~Thorlacius, \emph{{Zero mode dynamics of string webs}},
  \href{https://doi.org/10.1103/PhysRevD.63.126002}{\emph{Phys. Rev.}
  {\bfseries D63} (2001) 126002},
  [\href{https://arxiv.org/abs/hep-th/0101080}{{\ttfamily hep-th/0101080}}].

\bibitem{Lunin:2001jy}
O.~Lunin and S.~D. Mathur, \emph{{AdS / CFT duality and the black hole
  information paradox}},
  \href{https://doi.org/10.1016/S0550-3213(01)00620-4}{\emph{Nucl. Phys.}
  {\bfseries B623} (2002) 342--394},
  [\href{https://arxiv.org/abs/hep-th/0109154}{{\ttfamily hep-th/0109154}}].

\bibitem{Giusto:2004id}
S.~Giusto, S.~D. Mathur and A.~Saxena, \emph{{Dual geometries for a set of
  3-charge microstates}},
  \href{https://doi.org/10.1016/j.nuclphysb.2004.09.001}{\emph{Nucl. Phys.}
  {\bfseries B701} (2004) 357--379},
  [\href{https://arxiv.org/abs/hep-th/0405017}{{\ttfamily hep-th/0405017}}].

\bibitem{Bena:2015bea}
I.~Bena, S.~Giusto, R.~Russo, M.~Shigemori and N.~P. Warner, \emph{{Habemus
  Superstratum! A constructive proof of the existence of superstrata}},
  \href{https://doi.org/10.1007/JHEP05(2015)110}{\emph{JHEP} {\bfseries 05}
  (2015) 110}, [\href{https://arxiv.org/abs/1503.01463}{{\ttfamily
  1503.01463}}].

\bibitem{Bena:2016ypk}
I.~Bena, S.~Giusto, E.~J. Martinec, R.~Russo, M.~Shigemori, D.~Turton et~al.,
  \emph{{Smooth horizonless geometries deep inside the black-hole regime}},
  \href{https://doi.org/10.1103/PhysRevLett.117.201601}{\emph{Phys. Rev. Lett.}
  {\bfseries 117} (2016) 201601},
  [\href{https://arxiv.org/abs/1607.03908}{{\ttfamily 1607.03908}}].

\bibitem{Bena:2017xbt}
I.~Bena, S.~Giusto, E.~J. Martinec, R.~Russo, M.~Shigemori, D.~Turton et~al.,
  \emph{{Asymptotically-flat supergravity solutions deep inside the black-hole
  regime}}, \href{https://doi.org/10.1007/JHEP02(2018)014}{\emph{JHEP}
  {\bfseries 02} (2018) 014},
  [\href{https://arxiv.org/abs/1711.10474}{{\ttfamily 1711.10474}}].

\bibitem{Bena:2018bbd}
I.~Bena, P.~Heidmann and D.~Turton, \emph{{AdS$_{2}$ holography: mind the
  cap}}, \href{https://doi.org/10.1007/JHEP12(2018)028}{\emph{JHEP} {\bfseries
  12} (2018) 028}, [\href{https://arxiv.org/abs/1806.02834}{{\ttfamily
  1806.02834}}].

\end{thebibliography}\endgroup



\providecommand{\href}[2]{#2}\begingroup\raggedright\endgroup
\end{document}